\documentclass[fleqn,usenatbib,usedcolumn]{mnras}
\usepackage[british]{babel}             
\usepackage{txfonts}                  
\let\la=\lesssim 
\usepackage[T1]{fontenc}                
\usepackage{graphicx}                   
\usepackage{rotating}
\usepackage{dcolumn}
\newcolumntype{d}{D{.}{.}{-1}}
\def\dhead#1{\multicolumn{1}{c}{#1}}
\def\twolines#1#2{$\kern-6pt\Big\{ {\textrm{#1\hfill}\atop\textrm{#2\hfill}}$}

\usepackage{multirow}
\usepackage{hyperref}
\graphicspath{{.}{./FIGS/}}

\hypersetup{pdfauthor={I. H. Whittam, J. M. Riley, D. A. Green and M. J. Jarvis},
               pdftitle={GMRT 610-MHz observations of the faint radio source population},
               pdfkeywords={galaxies: active, radio continuum: galaxies, catalogues, surveys},
               bookmarksnumbered=true}
\hypersetup{colorlinks=true,
            linkcolor=blue,
            citecolor=blue,
            filecolor=blue,
            urlcolor=blue}
\volume{{\rm in prep.}}
\title[GMRT 610-MHz observations of the faint radio source population]{GMRT 610-MHz observations of the faint radio source population -- and what these tell us about the higher-radio-frequency sky}

\author[I.~H.~Whittam et al.]{I.~H.~Whittam$^{1}$\thanks{email:
\texttt{iwhittam@uwc.ac.za}}, D.~A.~Green$^2$, M. J. Jarvis$^{1,3}$ and J.~M.~Riley$^2$\\
   $^{1}$Department of Physics and Astronomy, University of the Western Cape, Robert Sobukwe Road, Bellville 7535, South Africa\\
   $^{2}$Astrophysics Group, Cavendish Laboratory, 19 J.~J.~Thomson Avenue, Cambridge CB3 0HE\\
   $^{3}$Astrophysics, University of Oxford, Denys Wilkinson Building, Keble Road, Oxford, OX1 3RH}

\date{Accepted ---; received ---; in original form ---}

\pagerange{\pageref{firstpage}--\pageref{lastpage}}

\pubyear{2016}
\begin{document}

\label{firstpage}

\maketitle

\begin{abstract}
We present 610-MHz Giant Metrewave Radio Telescope observations of 0.84~deg$^2$ of the AMI001 field (centred on $00^{\rm h} 23^{\rm m} 10^{\rm s}$, $+31^{\circ} 53'$) with an r.m.s.\ noise of 18 $\muup$Jy beam$^{-1}$ in the centre of the field. 955 sources are detected, and 814 are included in the source count analysis. The source counts from these observations are consistent with previous work. We have used these data to study the spectral index distribution of a sample of sources selected at 15.7 GHz from the recent deep extension to the Tenth Cambridge (10C) survey. The median spectral index, $\alpha$, (where $S \propto \nu^{-\alpha}$) between $0.08 < S_{15.7~\rm GHz} / \rm mJy < 0.2$ is $0.32 \pm 0.14$, showing that star-forming galaxies, which have much steeper spectra, are not contributing significantly to this population. This is in contrast to several models, but in agreement with the results from the 10C ultra-deep source counts; the high-frequency sky therefore continues to be dominated by radio galaxies down to $S_{15.7~\rm GHz} = 0.1$~mJy.
\end{abstract}


\section{Introduction}\label{section:intro}

The sub-mJy radio source population has been the subject of considerable study over the last decade, particularly at 1.4~GHz, with observations pushing down to the $\muup$Jy level (e.g.\ \citealt{2006MNRAS.371..963B,2008AJ....136.1889O,2012ApJ...758...23C}). Studies of the differential source counts derived from these observations allow us to probe how the different galaxy populations contribute to the total number of objects in the universe, and how the luminosity functions of these objects change with cosmic epoch. It is now widely accepted that the inflection in the differential source counts observed at $S_{1.4~\rm GHz} \sim 1$~mJy is due to the emergence of star-forming galaxies and radio-quiet AGN, which begin to contribute significantly to the radio sky below $S_{1.4~\rm GHz} \sim 1$~mJy (e.g.\ \citealt{2004NewAR..48.1173J,2008MNRAS.386.1695S,2008ApJS..177...14S,2009ApJ...694..235P,2015MNRAS.448.2665W}). An in-depth review of radio source counts and their implications is given by \citet{2010A&ARv..18....1D}. 

There is significant scatter in different measurements of the differential source counts at 1.4~GHz, many of which do not agree with each other within their respective errors (see e.g.\ \citealt{2005AJ....130.1373H,2006MNRAS.371..963B,2007ASPC..380..189C,2008AJ....136.1889O}). There has been significant debate about the origin of this scatter, but \citet{2013MNRAS.432.2625H} showed that sample variance alone cannot account for the differences found, pointing towards instrumental effects.

Interest in the faint radio sky was heightened by the Absolute Radiometer for Cosmology, Astrophysics and Diffuse Emission (ARCADE2; \citealt{2011ApJ...734....5F}) balloon experiment, which showed that there was a significant excess in the sky brightness temperature at 3~GHz which cannot be explained by current models \citep{2011ApJ...734....6S}. If this result is genuine, it suggests that there is a population of unknown radio sources at the $\muup$Jy or nJy level. 

The extragalactic source population at higher frequencies ($\gtrsim 10$~GHz) has been much less widely studied, mostly due to the increased telescope time required to carry out a survey to an equivalent depth over a significant area at higher frequencies. One exception is the Tenth Cambridge (10C; \citealt{2011MNRAS.415.2699F,2011MNRAS.415.2708D}) survey made with the Arcminute Microkelvin Imager (AMI) Large Array \citep{2008MNRAS.391.1545Z}, in the deep part of which 12~deg$^2$ were observed at 15.7~GHz to a completeness limit of 0.5~mJy across ten different fields. Study of the 10C sources \citep{2011MNRAS.415.2708D,2013MNRAS.429.2080W} has shown that their properties do not match those predicted by several leading models (e.g.\ \citealt{2005A&A...431..893D,2008MNRAS.388.1335W}), demonstrating the need to study this population rather than rely on extrapolations from lower frequencies. Specifically, \citet{2013MNRAS.429.2080W} showed that there is a population of faint, flat-spectrum sources present in the 10C survey which is not predicted by the models. Further study has shown that these sources are the cores of faint radio galaxies (\citealt{2015MNRAS.453.4244W}; \citealt{2016arXiv160703709W}), which are both more numerous and have flatter spectra than expected.

A recent deeper continuation of the 10C survey (10C ultra-deep, \citealt{2016MNRAS.457.1496W}) has extended the 10C survey to 0.1~mJy in two fields.  
Due to telescope scheduling, the deeper of these 15.7~GHz surveys is in the AMI001 field, a region of the sky with very little in the way of complementary data as the field was chosen purely on radio grounds. 
In order to investigate this sample further and constrain the proportions of different source types which contribute to this population it is vital that we have observations at other frequencies; we have therefore used the Giant Metrewave Radio Telescope (GMRT) to observe this field at 610 MHz. This paper describes these 610-MHz observations, and their implications for the nature of the higher-radio-frequency source population. 

The first part of this paper describes deep 610-MHz GMRT observations of the AMI001 field; in Section~\ref{section:obs} the observations and data reduction are explained, the source extraction is described in Section~\ref{section:catalogue} and the source counts are presented in Section~\ref{section:counts}. We then show how these observations can provide vital information about the higher-frequency population; the spectral index distribution of a 15.7-GHz-selected sample is presented in Section~\ref{section:alpha} and the conclusions are given in Section~\ref{section:conclusions}.

\section{Observations and data reduction}\label{section:obs}

\begin{figure}
\centerline{\includegraphics[width=\columnwidth]{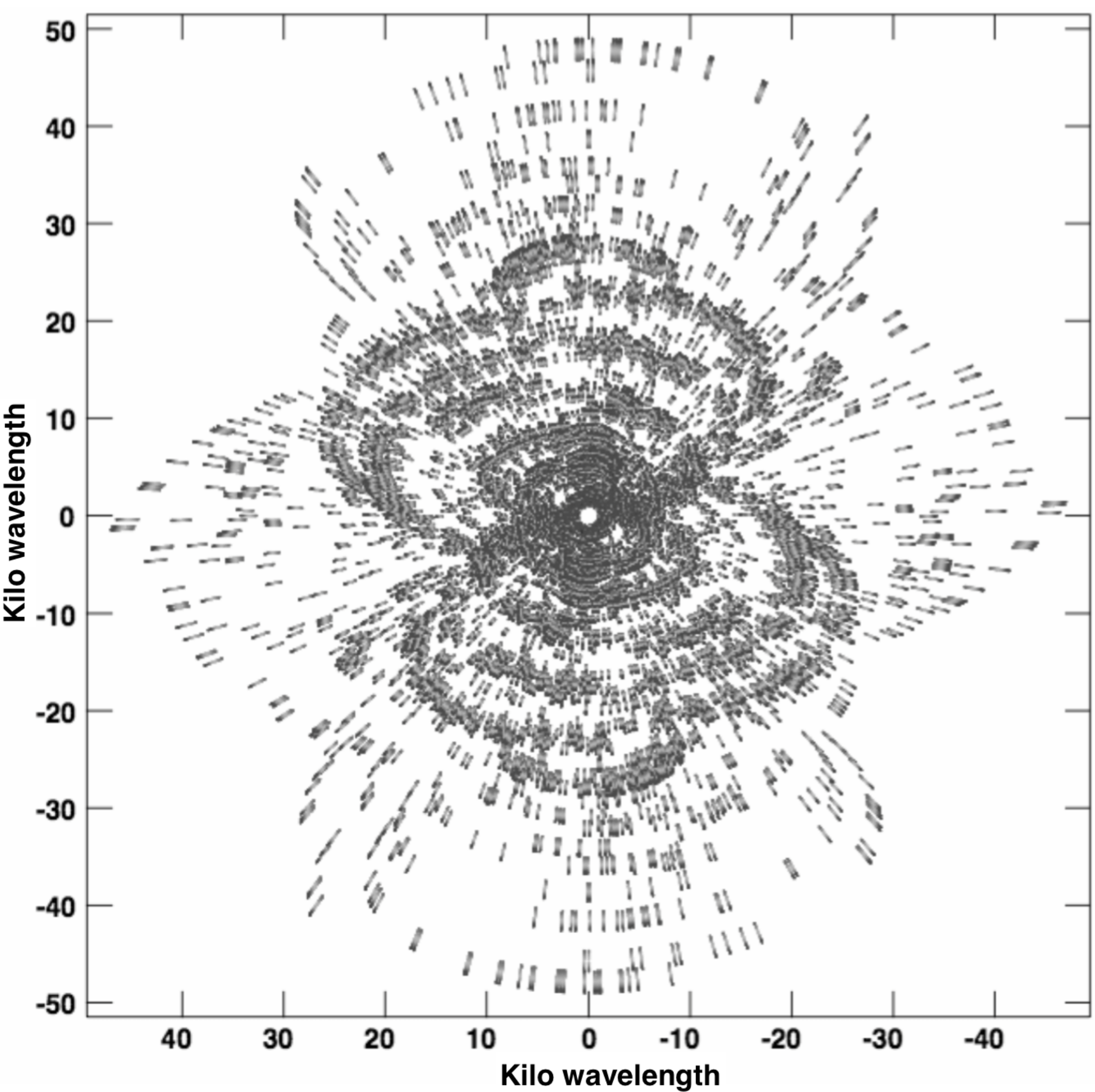}}
\caption{The $(u,v)$ coverage. Baselines shorter than 1~k$\lambda$ were not included in the imaging and are omitted from this plot.}\label{fig:uv}
\end{figure}

The AMI001 field was observed with a single pointing centred on $00^{\rm h} 23^{\rm m} 10$, $+31^{\circ} 53'$ (J2000 coordinates, used throughout) on 7~January 2012 with the Giant Metrewave Radio Telescope (GMRT) operating at 610~MHz. The pointing centre was chosen to be away from bright sources so as to minimise dynamic range issues. At this frequency the GMRT has a primary beam of 44 arcmin (FWHM) and an angular resolution of $\approx 7$ arcsec. The field was observed for 11 hours, including calibration, with a bandwidth of 32 MHz which was split into 256 spectral channels.  The radio sources 3C286 and 3C48 were observed at the beginning, in the middle and at the end of the run as primary flux density calibrators. The nearby radio source J0029+346 was observed for 5 minutes every 30 minutes as a secondary flux density and phase calibrator. The total time on source was 400~minutes. The resulting $(u,v)$ coverage is shown in Fig.~\ref{fig:uv}.

\begin{figure*}
\centerline{\includegraphics[width=16cm]{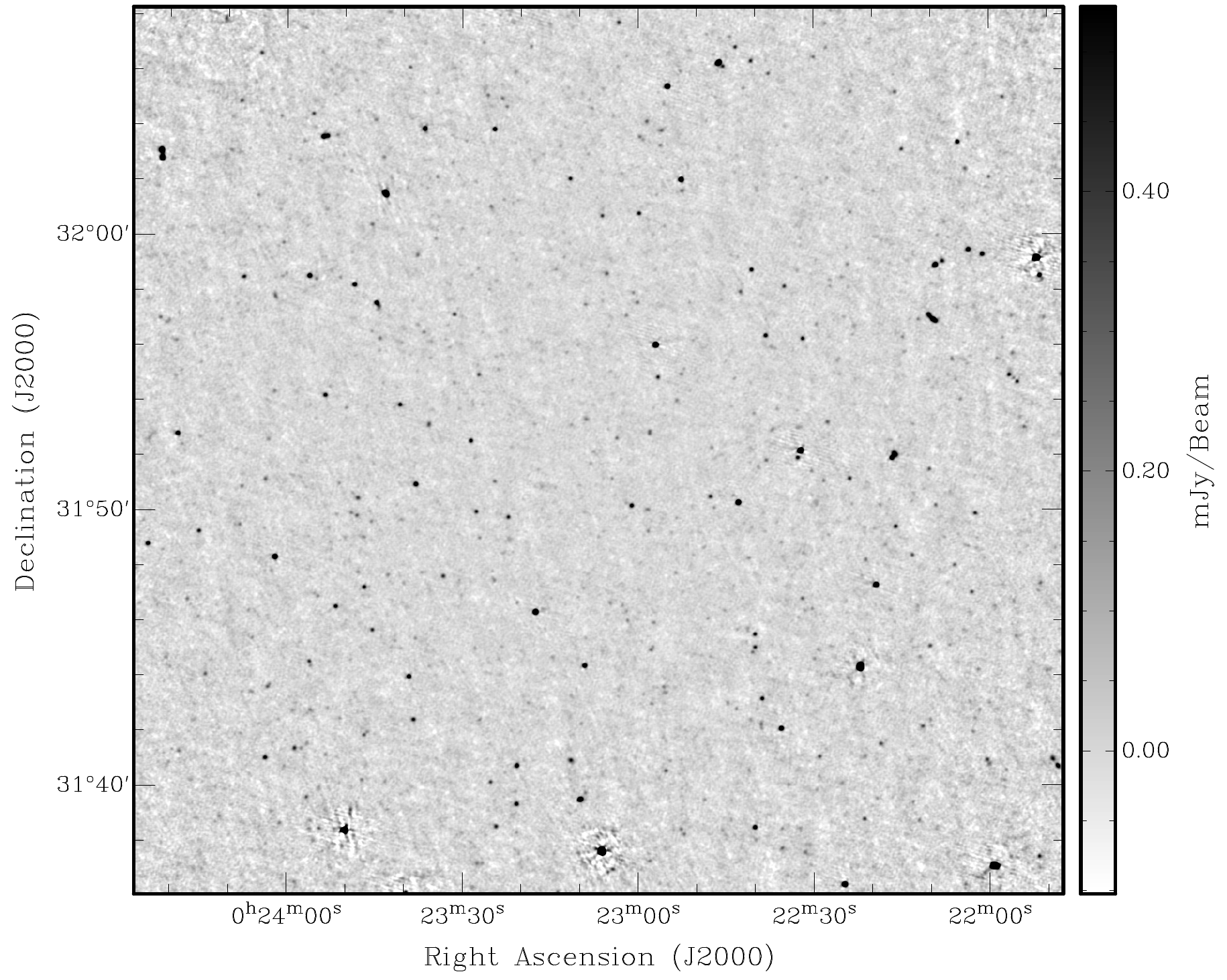}}
\caption{The central $40 \times 30$ arcmin of the 610-MHz GMRT image after primary beam correction. The synthesised beam size is $7.6 \times 7.3$ arcsec and the primary beam has a full-width half maximum of 44~arcmin.}\label{fig:map}
\end{figure*}

The data reduction was performed using the Astronomical Image Processing Software (\textsc{aips})\footnote{http://www.aips.nrao.edu/} package. The \textsc{aips} task \textsc{setjy} was used to calculate 610-MHz flux densities of 3C286 and 3C48 as 21.1 and 29.4 Jy respectively \citep{2013ApJS..204...19P}. Standard \textsc{aips} tasks were used to flag bad channels, baselines and antennas suffering from interference. A bandpass correction was then applied using the primary calibrators. Five central channels which were relatively free from interference were averaged to create a pseudo-continuum channel and an antenna-based phase and amplitude calibration was created using observations of J0029+346. This calibration was then applied to the full 256 channel dataset. The dataset was then compressed into 23 channels, each containing 10 original channels (the first and last few spectral channels were omitted as they tend to have large bandpass corrections). Further flagging was then performed on the 23 channel dataset. One antenna required re-weighting as it was noticed that the weights were discrepant. This was carried out using a custom \textsc{aips} task.

If we had imaged the whole field at once the non-planar nature of the sky would have caused the introduction of phase errors. To avoid this, the field was split into 31 smaller facets which were imaged separately with different phase centres, and then recombined. An additional six small facets around bright sources outside the imaging area were also included to account for contributions from these sources.

\begin{figure}
\centerline{\includegraphics[width=\columnwidth]{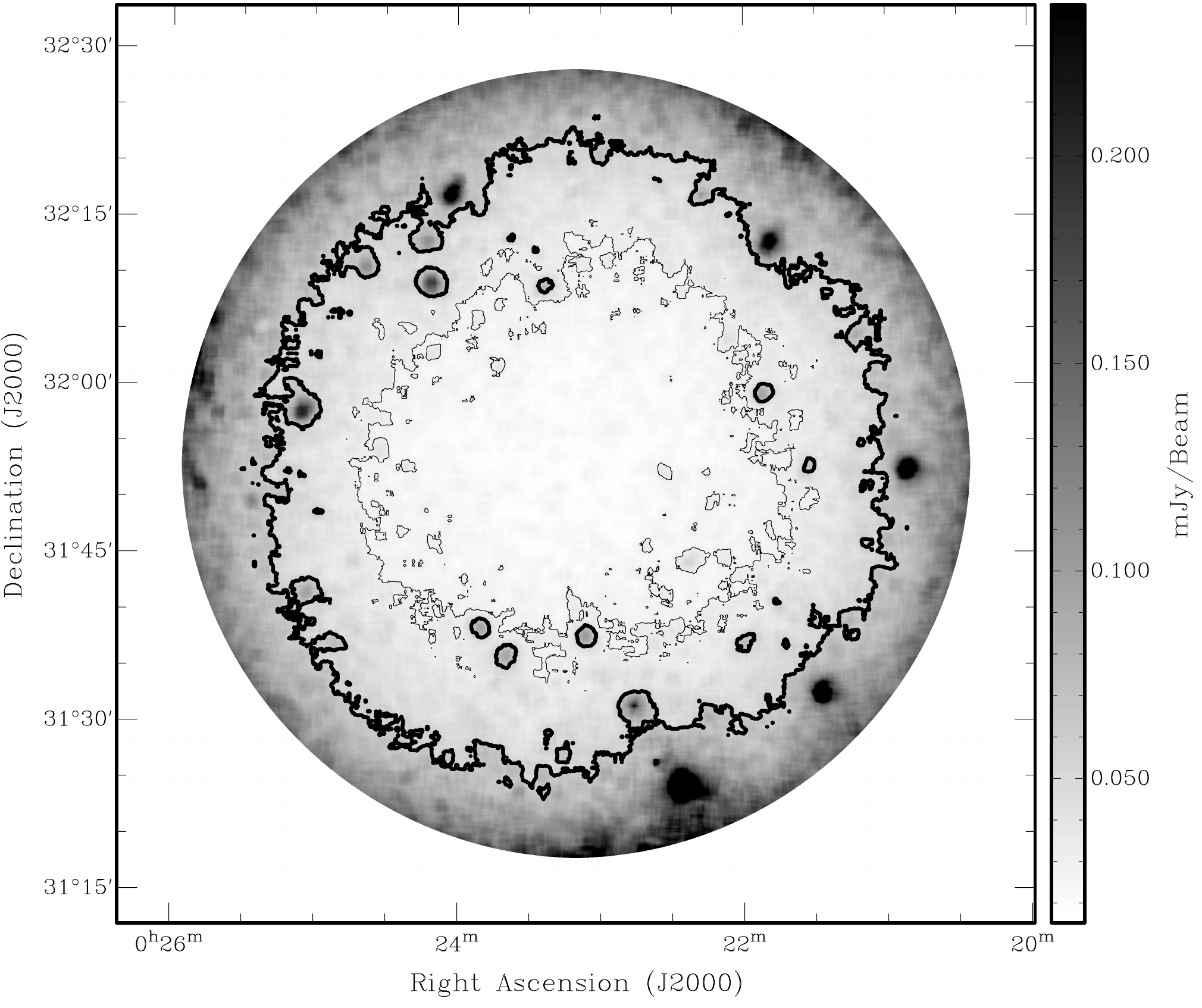}}
\caption{Noise map of the 610-MHz GMRT observations of the AMI001 field after primary beam correction. Contours are at 30 and 60~$\muup$Jy.}\label{fig:noise}
\end{figure}

The lack of bright sources in the field meant that care had to be taken when performing the self-calibration (the brightest source in the field has a flux density of 8~mJy). For the self-calibration a lower resolution image, with beam size 8.3~arcsec, was created. This lower-resolution image then went through three iterations of phase self-calibration with 15-, 5- and 2-minute solution intervals before a final iteration of phase and amplitude self-calibration with 15-minute intervals. The overall amplitude gain was held constant to ensure the flux density of the sources was not altered. A final, full resolution image was then created with a synthesised beam size of $7.6 \times 7.3~\rm{arcsec}$, PA $-7.5 ^{\circ}$ and with a pixel size of 1.4~arcsec to ensure the beam was oversampled. Natural weighting was used (\textsc{imagr} parameter \textsc{robust} = 5); baselines shorter than 1~k$\lambda$ were omitted from the imaging as the GMRT has a large number of short baselines which would otherwise dominate the beam shape, and are also prone to interference. (This corresponds to scales larger than 1~arcmin; few, if any, sources in the field are expected to have emission on this scale.)  The image was cut off at the point where the primary beam fell below 10 per cent. The r.m.s.\ in the centre of the field is $18~ \muup \rm{Jy}~\rm{beam}^{-1}$ before primary beam correction. 
The central part of the image is shown in Fig. \ref{fig:map} and the noise map of the field is shown in Fig.~\ref{fig:noise}. The dynamic range (ratio of peak flux density to lowest noise in the map before primary beam correction) in the map is $\approx 8000$; an alternative very conservative definition of dynamic range is the ratio of the peak flux density to the largest known artefact nearby, estimated using the largest `negative peak', which gives a dynamic range of 50.

\section{Source catalogue}\label{section:catalogue}

\subsection{Source finding}\label{section:source_finding}

The source fitting was carried out using the \textsc{source\_find} software, which is described in \citet{2011MNRAS.415.2699F} and summarised briefly here. A noise map is produced by the software as follows; at each pixel position, the noise is taken as the r.m.s.\ inside a square centred on the pixel with a width of approximately ten times the synthesized beam. The software then uses the image along with the noise map to identify peaks in the map above a given signal to noise value $\gamma$ (here, $\gamma = 5$). All peaks with flux densities greater than $0.6 \gamma \sigma$ (where $\sigma$ is the local noise value) are identified initially to ensure that all peaks greater than $\gamma \sigma$ after interpolation between the pixels are included. The position and value of each peak (RA$_{\rm pk}$, Dec$_{\rm pk}$ and $S_{\rm pk}$) are then found by interpolation between the pixels, and any peaks less than $\gamma \sigma$ are discarded. The error on the peak flux density ($\Delta S_{\rm pk}$) is taken to be the thermal noise error combined in quadrature with a conservative ten per cent calibration error, $\Delta S_{\rm pk} = \sqrt{\sigma^2 + (0.1S_{\rm pk})^2}$. The integration area, consisting of contiguous pixels down to a lowest contour value of 2.5$\sigma$, is then calculated for each component. Sources are classified as being part of a `group' if more than one peak is found inside the same integration area.

The integrated flux density, position and angular size ($S_{\rm int}$, RA$_{\rm int}$, Dec$_{\rm int}$ and $e_{\rm maj}$) are estimated for each component automatically using the \textsc{aips} task \textsc{jmfit}, which fits a 2D Gaussian to each component. The error on the integrated flux density ($\Delta S_{\rm int}$) is estimated as the error due to thermal noise (estimated by \textsc{jmfit}) combined in quadrature with a conservative ten per cent calibration error, $\Delta S_{\rm int} = \sqrt{\sigma^2 + (0.1S_{\rm int})^2}$.

A source is considered to be extended if the major axis of the deconvolved Gaussian ($e_{\rm maj}$) is larger than a critical value $e_{\rm crit}$ (see \citealt{2011MNRAS.415.2699F}), where
\begin{equation}
e_{\rm crit} =  \left\{ 
                  \begin{array}{l l}
                  3.0\, b_{\rm maj}\, \rho^{-1/2}  & \textrm{if} ~3.0\, b_{\rm maj} \, \rho^{-1/2} > 6.0~\textrm{arcsec},\\
                  6.0~\rm arcsec            & \textrm{otherwise},\\
                  \end{array}\right.
\label{eqn:ecrit}
\end{equation}
where $b_{\rm maj}$ is the major axis of the restoring beam and $\rho = S_{\rm pk} / \sigma$ (i.e.\ the signal-to-noise ratio). Sources with $e_{\rm maj} > e_{\rm crit}$ were classified as extended (flag E), otherwise the source was considered point-like (flag P).

\subsection{The source catalogue}\label{section:catalogue}

There are 955 components in the catalogue, 151 of which are flagged as being part of a group. The full source catalogue is available online\footnote{http://vizier.u-strasbg.fr/viz-bin/VizieR}, with a sample of ten rows shown in Table~\ref{tab:cat_sample}, where the columns are:

\noindent (1) -- source name;

\noindent (2) -- group designation (group name followed by number of components in group in brackets);

\noindent (3) and (4) -- peak right ascension and declination (J2000);

\noindent (5) -- peak flux density (mJy);

\noindent (6) -- error on the peak flux density (mJy);

\noindent (7) -- integrated flux density (mJy);

\noindent (8) -- error on the integrated flux density (mJy);

\noindent (9) and (10) -- deconvolved source major and minor axes (arcsec);

\noindent (11) -- deconvolved position angle, measured from North through to East ($^{\circ}$);

\noindent (12) -- local r.m.s.\ noise (mJy beam$^{-1}$);

\noindent (13) -- source type (E = extended, P = point-like), as classified using equation~\ref{eqn:ecrit}.

\begin{figure}
\centerline{\includegraphics[width=\columnwidth]{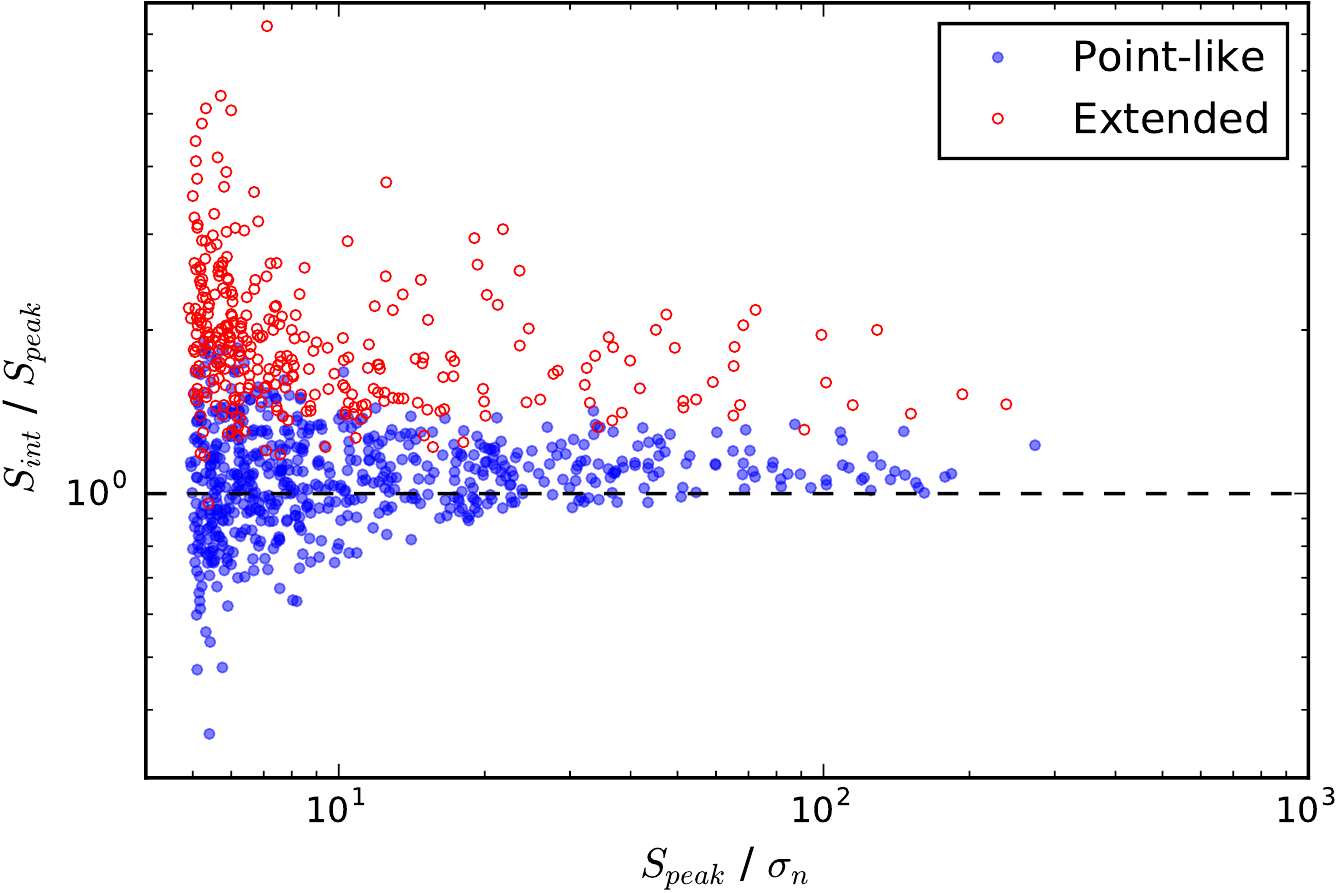}}
\caption{Ratio of the integrated flux density to peak flux density as a function of signal-to-noise ratio ($S_{\rm peak} / \sigma$). Sources which are classified as point-like and extended during the source-fitting procedure (see Section~\ref{section:source_finding}) are shown separately. The dashed line is at $S_{\rm int} /  S_{\rm peak} = 1$.}\label{fig:Speak}
\end{figure}

\begin{figure*}
\centerline{\hbox to 5.5cm{\hss (a)\hss}
            \quad
            \hbox to 5.5cm{\hss (b)\hss}
            \quad
            \hbox to 5.5cm{\hss (c)\hss}}
\centerline{\includegraphics[width=5.5cm]{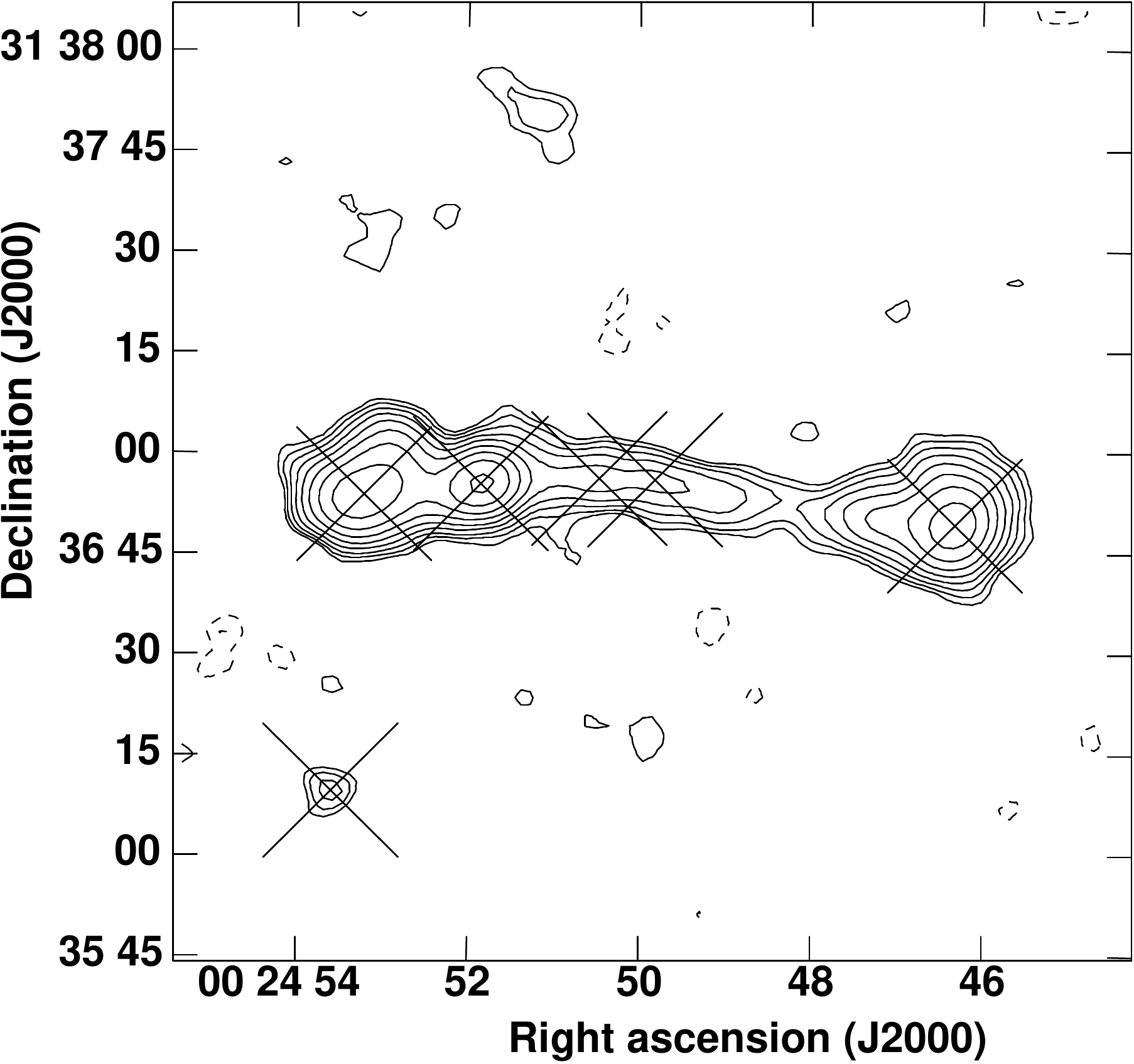}
            \quad
            \includegraphics[width=5.5cm]{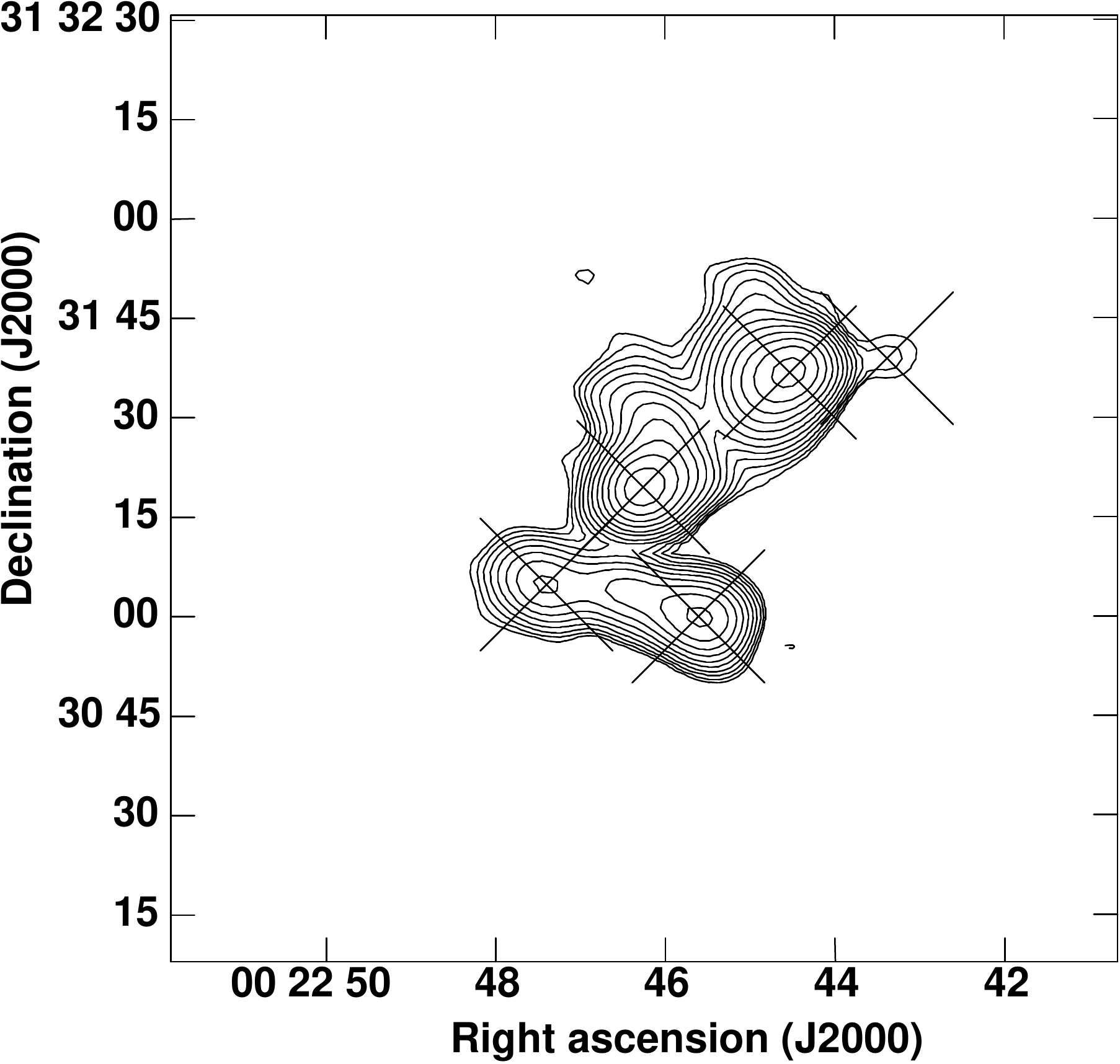}
            \quad
            \includegraphics[width=5.5cm]{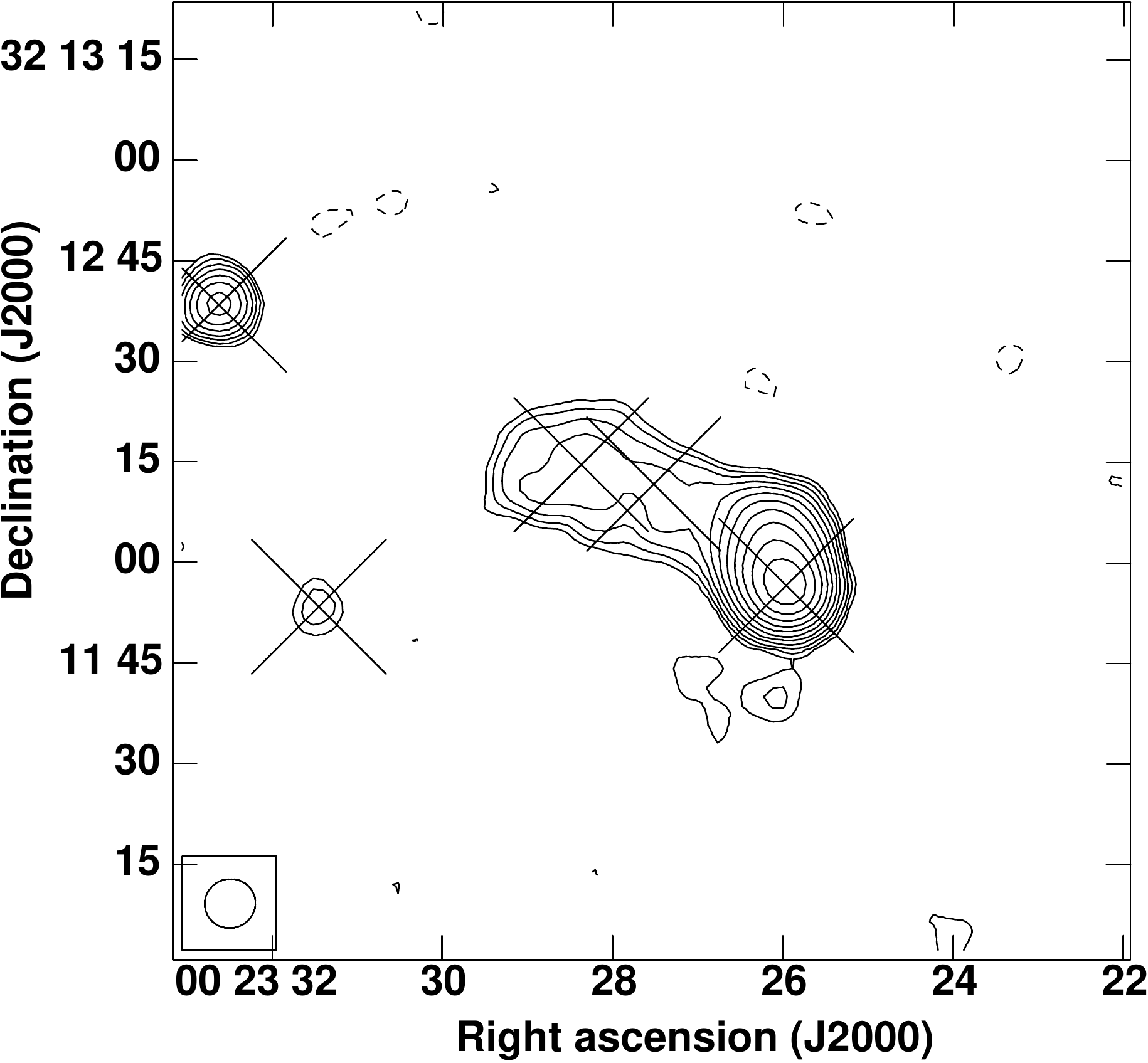}}
\caption{Three examples of extended sources found in the image. Crosses mark the peaks identified in the source fitting process. The contours are drawn at $(\pm 2 \sqrt{2^n}, n = 0, 1 ... 7) \times \sigma~\muup$Jy where $\sigma$ is the local r.m.s.\ noise. ($\sigma$ = 67 for (a), 199 for (b) and 49 for (c) ). }\label{fig:extended}
\end{figure*}

The ratio of the integrated to peak flux densities is shown as a function of signal-to-noise ratio in Fig.~\ref{fig:Speak}, with sources classified as point-like and extended shown separately. A total of 312 sources, approximately a third of those detected, are extended. Three examples of extended sources are shown in Fig.~\ref{fig:extended}.

\begin{table*}
\caption{A sample of ten rows from the source catalogue. See Section~\ref{section:catalogue} for a full description of the columns.}\label{tab:cat_sample}
\smallskip
\centering
\renewcommand{\tabcolsep}{1.5mm}
\begin{tabular}{ccccccdcccdcc}\hline
Source ID & Group & RA & Dec. & $S_{\rm peak}$ & $\Delta S_{\rm peak}$ & \dhead{$S_{\rm int}$} & $\Delta S_{\rm int}$ & $\theta_{\rm maj}$ & $\theta_{\rm min}$ & \dhead{PA}  & $\sigma$& Type\\
          &       &    &      & (mJy)          & (mJy)                 & \dhead{(mJy)}         & (mJy)                & (arcsec)           & (arcsec)      & \dhead{$(^{\circ})$} & (mJy beam$^{-1}$) & \\
(1) & (2) & (3) & (4) & (5) & (6) & \dhead{(7)} & (8) & (9) & (10) & \dhead{(11)} & (12) & (13)\\
\hline
002117+315202 & - & 00:21:17.06 & +31:52:02.03 & 0.553 & 0.077 & 0.85 & 0.173 & 8.7 & 6.1 & 128.8 & 0.054 & E\\
  002118+315907 & - & 00:21:18.85 & +31:59:07.48 & 0.332 & 0.055 & 0.599 & 0.158 & 11.8 & 7.1 & 157.8 & 0.044 & E\\
  002119+315639 & - & 00:21:19.05 & +31:56:39.15 & 2.931 & 0.297 & 3.439 & 0.345 & 4.1 & 1.4 & 141.1 & 0.049 & P\\
  002120+315900 & 002121+315914(02) & 00:21:20.80 & +31:59:00.22 & 1.134 & 0.123 & 2.136 & 0.245 & 8.8 & 5.0 & 24.2 & 0.048 & E\\
  002120+320438 & - & 00:21:20.36 & +32:04:38.66 & 0.44 & 0.073 & 0.65 & 0.178 & 12.3 & 3.1 & 72.4 & 0.058 & E\\
  002120+321025 & - & 00:21:20.98 & +32:10:25.07 & 1.948 & 0.21 & 4.418 & 0.45 & 10.4 & 5.2 & 7.7 & 0.079 & E\\
  002121+315914 & 002121+315914(02) & 00:21:21.63 & +31:59:14.94 & 1.319 & 0.138 & 2.144 & 0.229 & 6.8 & 4.6 & 70.5 & 0.041 & E\\
  002121+320307 & - & 00:21:21.62 & +32:03:07.89 & 0.271 & 0.053 & 0.308 & 0.104 & 7.7 & 0.0 & 129.7 & 0.046 & P\\
  002122+314613 & - & 00:21:22.89 & +31:46:13.02 & 0.75 & 0.086 & 0.907 & 0.122 & 4.5 & 2.6 & 106.5 & 0.042 & P\\
  002122+314901 & - & 00:21:22.38 & +31:49:01.35 & 0.256 & 0.052 & 0.263 & 0.097 & 8.2 & 0.0 & 178.4 & 0.045 & P\\
\hline\end{tabular}
\end{table*}

\subsection{Completeness}

\begin{figure}
\centerline{\includegraphics[width=\columnwidth]{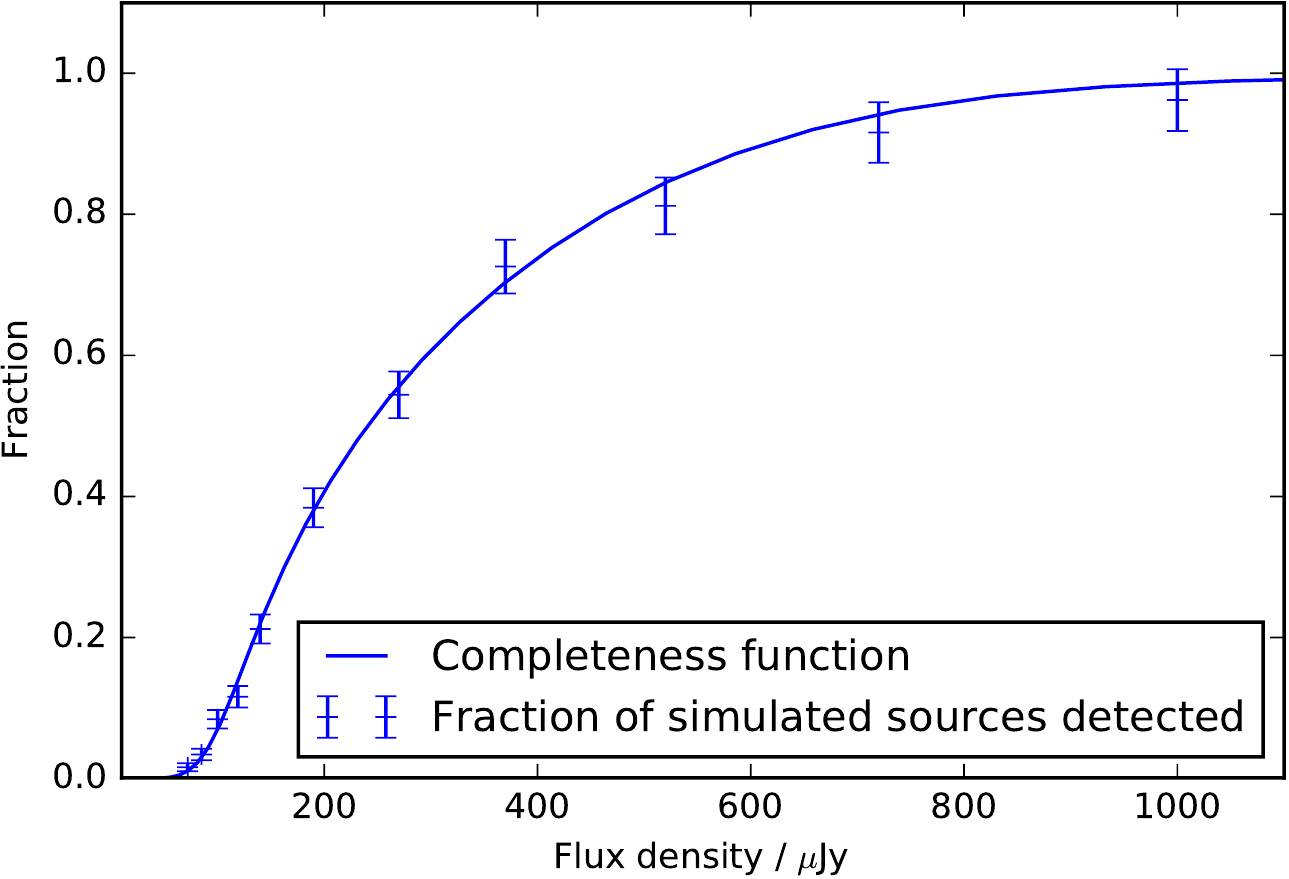}}
\caption{The completeness function (given by Equation~\ref{eqn:prob}, solid line) used to correct the source counts, and the fraction of simulated sources detected as a function of flux density in the completeness check (points). The errors plotted are Poisson errors. }\label{fig:completeness}
\end{figure}

The completeness of the catalogue was estimated by inserting 500 ideal point sources with equal flux density $S$ into the map in random positions. The sources were inserted in the image plane using the \textsc{aips} task \textsc{immod}.  The source finding was then carried out on this simulated map in the same way as described in Section~\ref{section:source_finding}. An inserted source was considered to be detected if there was a source in the output catalogue within 2.8~arcsec (2~pixels) of the inserted source position. This process was repeated several times with a range of flux densities, and the results are shown in Fig.~\ref{fig:completeness}. Assuming the noise is Gaussian, the theoretical completeness can be estimated from the noise map. The probability of detecting a source with true flux density $S_i$ located on a pixel with corresponding noise-map value of $\sigma$ is given by:
\begin{equation}
P(S_i > 5\sigma)  = \int_{5 \sigma}^\infty \frac{1}{\sqrt{2 \pi \sigma^2}} \exp \left( - \frac{(X - S_i)^2}{2 \sigma^2}\right) {\rm d}X,\label{eqn:prob}
\end{equation}
which is shown as the solid line in Figure~\ref{fig:completeness}. The results using the simulated sources agree with this curve within the errors.

\subsection{Reliability}

Assuming the noise is Gaussian, the probability of a false detection in the map can be calculated. The probability of drawing a value more than 5 standard deviations away from the mean in a Gaussian distribution is $1.5 \times 10^{-6}$. There are $1.4 \times 10^5$ synthesised beams in the map, so we expect less than one false detection in the image. 

Residual phase and amplitude errors mean that the noise is not purely Gaussian.  To investigate the effects that these may have on the false detection rate we have run the source-finding algorithm on an inverted image.  The rationale behind this is that noise fluctuations are equally likely to be positive or negative, but we expect signals from real sources to be positive only.  The source finding algorithm only detects positive peaks, so by inverting the image and running \textsc{source\_find} on the inverted map in exactly the same way as described in Section~\ref{section:source_finding} any sources detected result from noise on the map.  14 sources were detected in the inverted image (compared to 955 in the real image), giving a false detection rate of 1.5~per cent, which indicates, as expected, that the noise in the image is not purely Gaussian.

We expect source confusion to contribute approximately $3~\muup$Jy beam$^{-1}$ to the noise \citep{2012ApJ...758...23C}.

\subsection{GMRT primary beam}

The half-power beam width (HPBW) of the primary beam varies slightly between the GMRT antennas. To test the effect that this might have on the final images, two different images were made, one with a HPBW of 44.0~arcmin and one with 44.8~arcmin  (a value of 44.4~arcmin was used in the image described in Section~\ref{section:obs}). These values represent an over-estimate of the possible range of the mean HPBW, as the variations between the individual antennas is $\sim 2$~arcmin (N. G. Kantharia, private communication), resulting in an estimated error on the mean HBPW of 0.3~arcmin. The source fitting procedure described in Section~\ref{section:source_finding} was carried out on both of these images and the resulting two catalogues were compared. These two catalogues proved to be extremely similar: all the sources in one image were found in the other and vice versa and the ratio of the flux densities in the two images is $1.03 \pm 0.03$. Thus the possible small variations in the HPBW value used when applying the primary beam correction to the final image do not have a significant effect on the sources detected in the image.

\subsection{Astrometry}

There are two other catalogues available in this field: the 10C survey \citep{2011MNRAS.415.2708D,2011MNRAS.415.2699F,2016MNRAS.457.1496W} and the National Radio Astronomy Observatory (NRAO) Very Large Array (VLA) Sky Survey (NVSS; \citealt{1998AJ....115.1693C}). Both of these catalogues are relatively low resolution, with synthesised beam sizes of 30 and 45 arcsec respectively; this means they are not ideal for assessing the astrometric accuracy of our catalogue (this field is not covered by the Faint Images of the Radio Sky and Twenty-one cm).

Nevertheless we matched our source catalogue to both the 10C ultra-deep and NVSS catalogues (using a match radius of 15~arcsec in both cases) and found 132 matches to the 10C ultra-deep catalogue and 56 to NVSS (see Section~\ref{section:matching} for further details of the matching to the 10C catalogue). The mean offsets in right ascension and declination between our work and these two catalogues are shown in Table~\ref{tab:astrometry}. These offsets are all less than 1.5~arcsec, which is smaller than the positional accuracies of both the 10C and NVSS catalogues. We therefore believe our astrometry is accurate at the $\sim$arcsec level.

\begin{table}
\caption{Mean positional offsets between this work (GMRT) and the NVSS and 10C catalogues.}\label{tab:astrometry}
\smallskip
\centering
\begin{tabular}{l|cc}
 & RA offset (arcsec) & Dec offset (arcsec)\\\hline
 NVSS - GMRT & $-0.8 \pm 0.4$ & $0.2 \pm 0.5$\\
 10C - GMRT  & $-0.6 \pm 0.3$ & $-1.2 \pm 0.3$\\\hline
\end{tabular}
\end{table}

\section{Source counts}\label{section:counts}

To reduce uncertainty when calculating the source counts, only sources within the 20 per cent power point of the primary beam are included in the catalogue used. This means that 62 sources towards the edge of the map included in the catalogue described in Section~\ref{section:catalogue} are not included in the catalogue used to calculate the source counts. If a source is classified as extended using the criteria described in Section~\ref{section:source_finding}, the integrated flux density is used when calculating the source counts. If, however, a source is point-like the peak flux density is used, as this provides a better measure of the flux density of unresolved sources.

\subsection{Sources with multiple components}

A fraction of the detected radio sources are resolved into multiple components, which needs to be considered when calculating the source counts. Components of a multiple source are listed as separate entries in the source catalogue produced by \textsc{source\_find}, but are flagged as being part of a group if more than one peak lies inside the same integration area (see Section~\ref{section:source_finding}).  A total of 145 sources form 66 groups in this catalogue. To calculate the source counts we `collapsed' the flux densities of these multiple sources by summing the flux densities of all the sources listed as being part of any one group and listing that as a single entry in the source catalogue used for calculating the source counts. We therefore have 814 sources in the final source count catalogue. 

As a check, the source count of the 893 individual components, and the catalogue after the groups have been collapsed, are compared. We find that the difference between the two counts is smaller than the size of the errorbars.

\subsection{Area around bright sources}\label{section:bright_sources}

Artefacts close to bright sources in radio images can lead to false detections in regions around bright sources. However, the region covered by these observations was chosen to avoid bright sources so there are very few bright sources in the catalogue. 
When investigating the effects of this, we consider the signal-to-noise ratio of a source, rather than the peak flux density, as the noise varies significantly across the image. \citet{2007MNRAS.376.1251G} studied the spatial density of sources found close to bright sources in their deep GMRT observations and found that false detections were an issue for sources with $S_{\rm peak} > 10$~mJy, which is equivalent to $S_{\rm peak} / \sigma = 200$. The lack of bright sources in our observations means that there are only two sources with $S_{\rm peak} / \sigma > 200$. Inspecting these two sources by eye we find no evidence for false detections due to artefacts close to either of them (there are no sources within 45 arcsec of either source). The effect of false detections close to bright sources is therefore not a significant issue in this catalogue so we make no attempts to correct for it.

\subsection{Calculating the source counts}\label{section:calc_counts}

As this observation consists of a single pointing, the noise varies across the field according to the primary beam shape (see Fig.~\ref{fig:noise}). In order to take this into account when calculating the source counts, we correct the contribution from each source by the area over which it could have been detected. Assuming the noise is Gaussian, which is the case away from bright sources, the probability of detecting a source with a peak flux density $S_i$ larger than 5$\sigma$ is given by Equation~\ref{eqn:prob}.

The variation in the noise across the map can be taken into account by averaging the probability of detecting a source (given by Equation~\ref{eqn:prob}) at each pixel position given the noise map. The resulting probability of detecting a source of a given peak flux density anywhere in the image is shown in Fig.~\ref{fig:completeness}. The contribution from each source is therefore corrected by the inverse of this fraction.
The source count in each flux density bin is therefore given by:
\begin{equation}
\frac{1}{A} \sum_{i=1}^{N} \frac{1}{P(S_i > 5\sigma)}
\end{equation}
where $A$ is the total area of the field, $N$ is the number of sources in the bin and $P(S_i > 5\sigma)$ is the probability of detecting a source with flux density $S_i$ in the field, given by Equation~\ref{eqn:prob}.

\subsection{Resolution bias}\label{section:bias}

When calculating source counts we require a catalogue which is complete in terms of integrated flux density, whereas sources are detected in terms of their peak flux density. This means that an extended source of a given integrated flux density is more likely to fall below the peak flux density detection limit than a point source with the same integrated flux density. This effect causes the number of sources to be underestimated, particularly near the detection limit of the survey. We correct for this effect in a similar way to \citet{2001A&A...365..392P} and \citet{2016MNRAS.tmp..926W}, as described below.

The following relation can be used to calculate the maximum angular size that a source can have and still be detected for a given total flux density ($S_{\rm int}$):
\begin{equation}
\frac{S_{\rm int}}{5\sigma} = \frac{\theta_{\rm min}\theta_{\rm maj}}{b_{\rm min}b_{\rm max}}
\end{equation}
where $\theta_{\rm min}$ and $\theta_{\rm maj}$ are the source FWHM axes, $b_{\rm min}$ and $b_{\rm max}$ are the synthesised beam FWHM axes and $5\sigma$ is the peak flux density detection limit (where $\sigma$ is the local r.m.s.\ noise in the image). We use this relation to calculate the maximum size $\Psi_{\rm max}$ that a source can have and still be detected, where $\Psi$ is the geometric mean of the source major and minor axes, for a given total flux density and local r.m.s.\ noise. As the noise varies significantly across the map the correction for resolution bias has been calculated for each source using the local r.m.s.\ noise and applied to the overall count in a similar way to the completeness correction described in Section~\ref{section:calc_counts}.

\begin{figure}
\centerline{\includegraphics[width=\columnwidth]{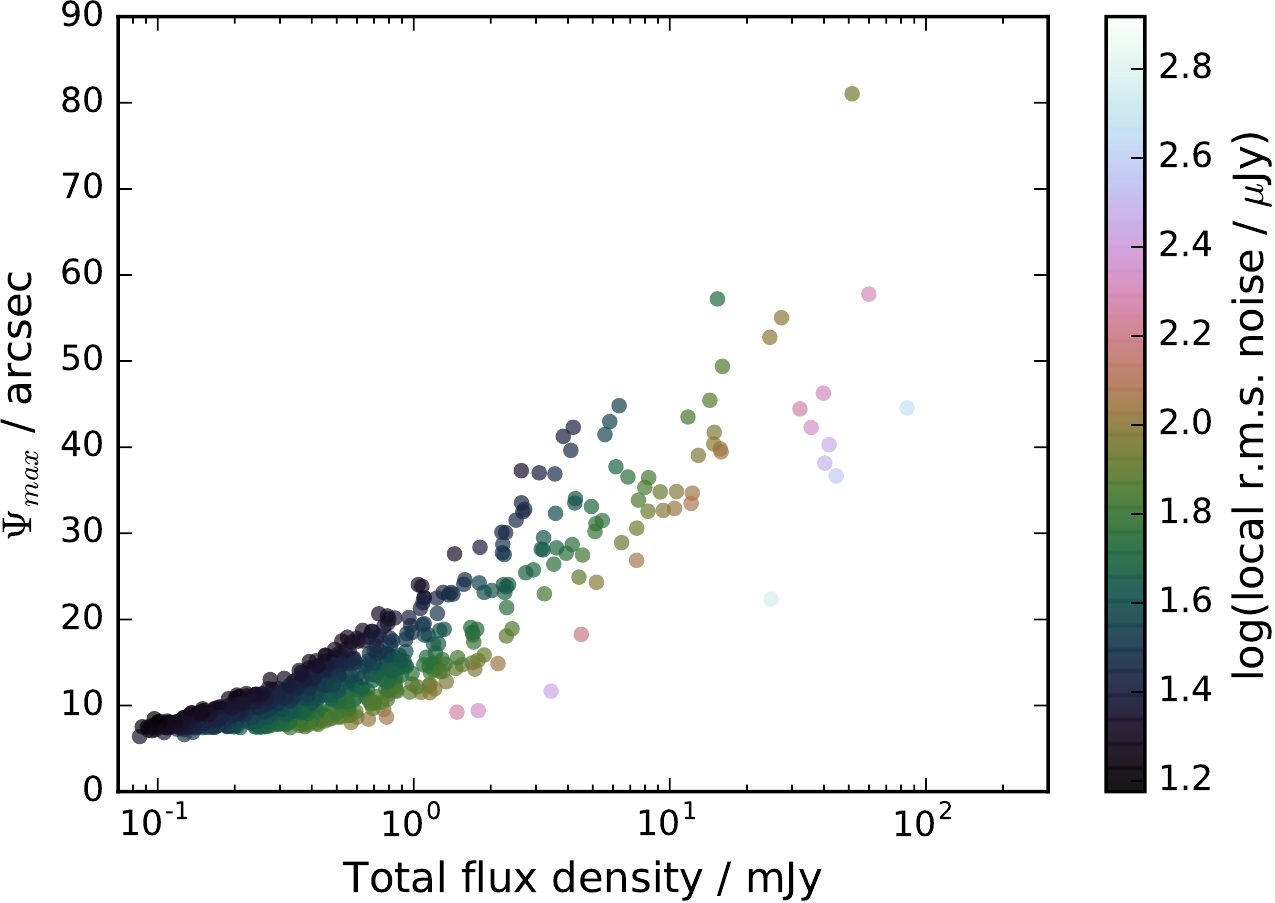}}
\caption{$\Psi_{\rm max}$ as a function of integrated flux density for each of the 814 sources in the source count catalogue. $\Psi_{\rm max}$ is the maximum angular size (geometric mean of the source major and minor axes) that a source can have and still be detectable (this depends on both the total flux density of each source and its local r.m.s. noise). Points are coloured according to their local r.m.s.\ noise using the cubehelix colour scheme \citep{2011BASI...39..289G}.}\label{fig:sizes}
\end{figure}

\begin{figure*}
\centerline{\includegraphics[width=\columnwidth]{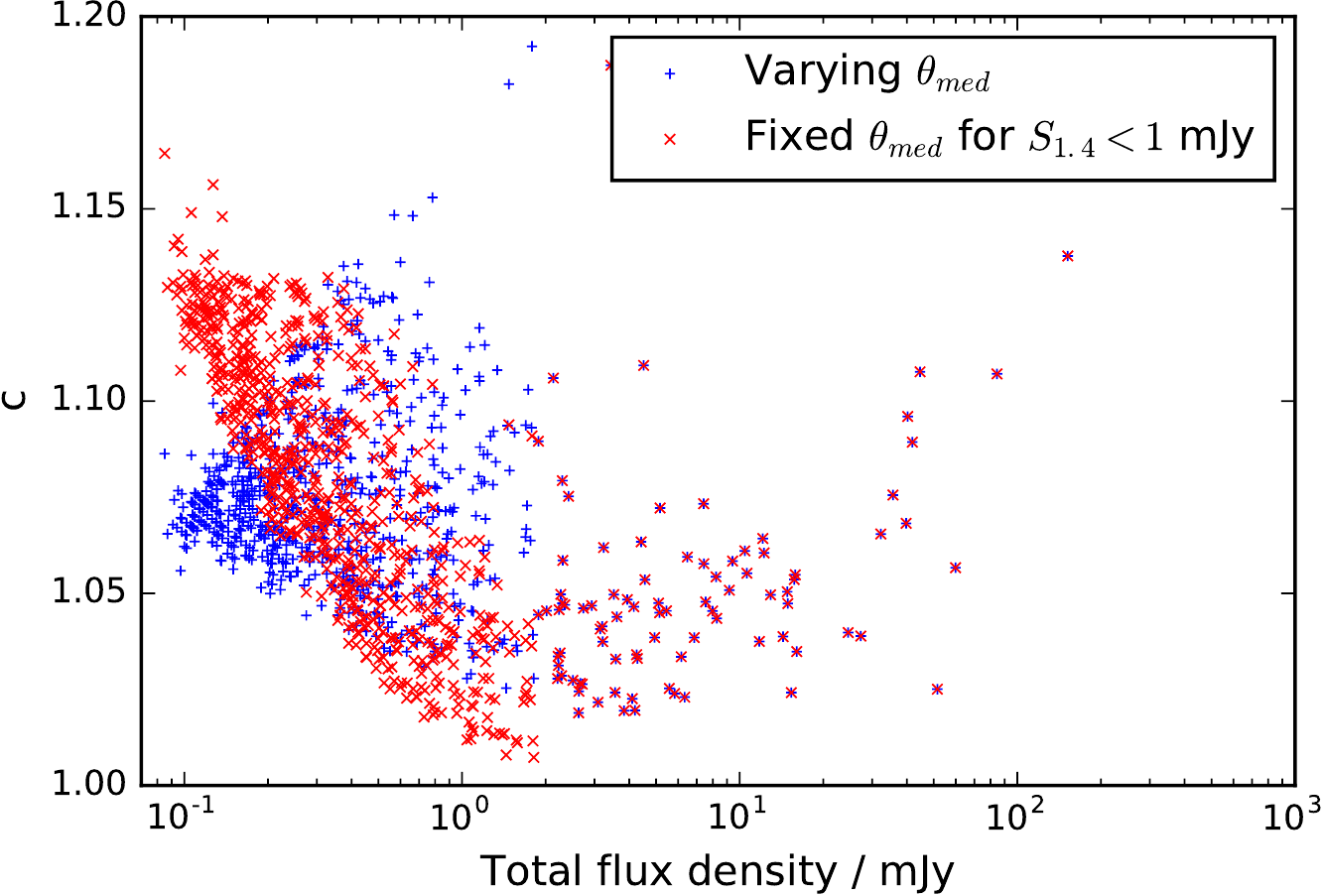}
            \quad
            \includegraphics[width=\columnwidth]{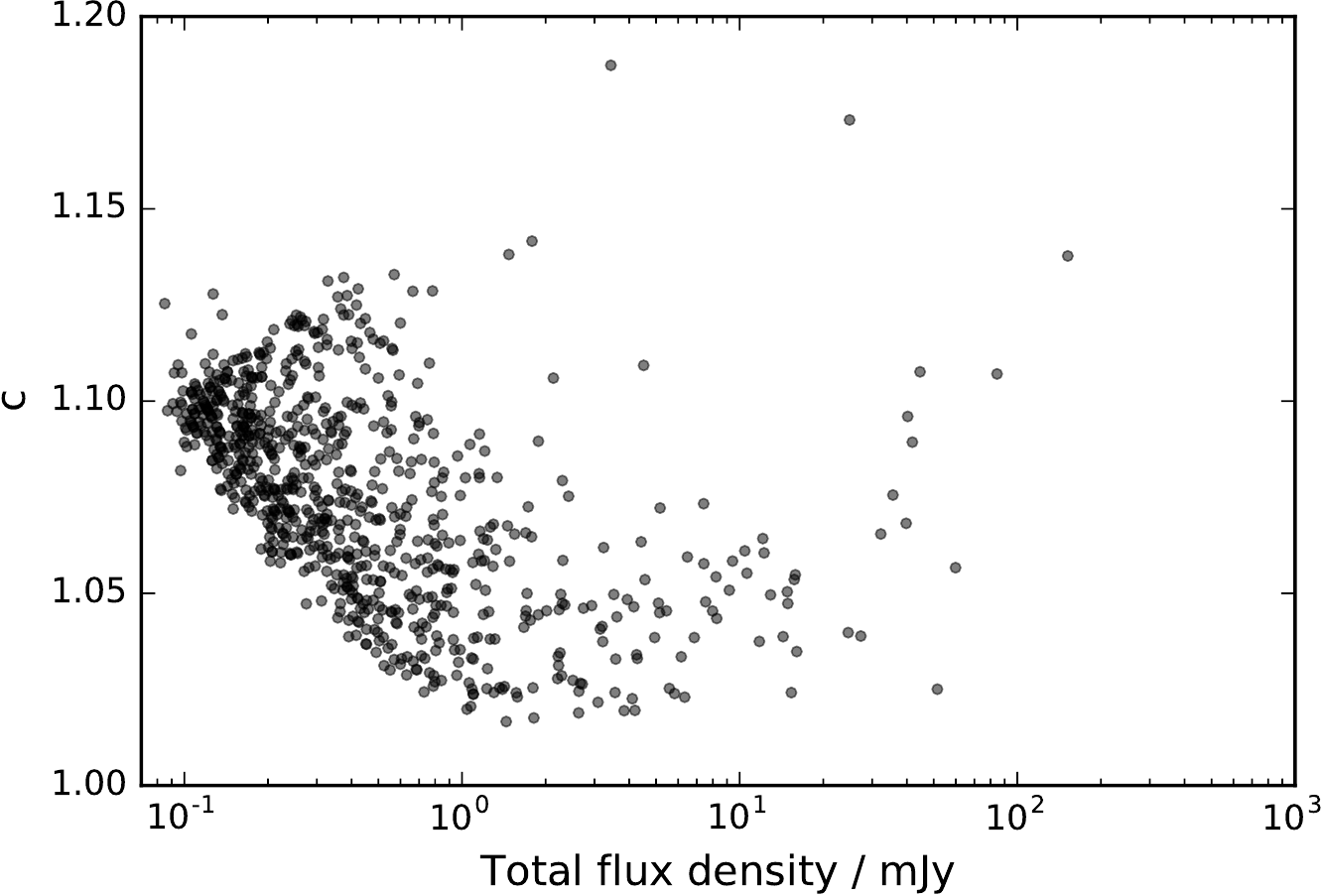}}
\caption{The resolution bias correction factor $c = 1/[1 - h(>\Psi_{\rm max})]$ as a function of flux density. The left-hand panel shows two different versions of the correction factor for each source; correction factors shown as $`+'$ (blue in the online version) are calculated using a variable $\theta_{\rm med} = 2(S_{1.4~\rm GHz})^{0.3}$~arcsec, while the values shown as $`\times'$ (red in the online version) used a constant $\theta_{\rm med} = 1.2$~arcsec for $S_{1.4~\rm GHz} < 1$~mJy (and the same variable $\theta_{\rm med}= 2(S_{1.4~\rm GHz})^{0.3}$~arcsec for sources brighter than 1~mJy). The right-hand panel shows the mean of these two correction factors, which is applied to the counts.}\label{figs:bias_correction}
\end{figure*}

The correction is calculated as follows. $\Psi_{\rm max}$ is calculated for each source in the catalogue and the distribution as a function of flux density is shown in Fig.~\ref{fig:sizes}. Given $\Psi_{\rm max}$, we calculate the fraction of sources expected to be larger than this value ($h(> \Psi_{\rm max})$), following \citet{1990ASPC...10..389W}, using:
\begin{equation}
h(> \Psi_{\rm max}) = {\rm exp}\left[- {\rm ln}(2) \left(\frac{\Psi_{\rm max}}{\theta_{\rm med}}\right)^{0.62}\right]
\end{equation}
where $\theta_{\rm med}$ is the median angular size. We use two different versions of $\theta_{\rm med}$ for comparison; the first is 
\begin{equation}
\theta_{\rm med} = 2(S_{1.4~\rm GHz})^{0.3}~{\rm arcsec}\label{eqn:thmed1}
\end{equation}
with $S_{1.4~\rm GHz}$ in mJy (flux densities are scaled from 610~MHz to 1.4~GHz using a spectral index of 0.75, which is appropriate for these frequencies), and the second used a constant size of 1.2~arcsec below 1~mJy (based on recent eMERLIN results, \citealt{2015aska.confE..23B}) as follows:
\begin{equation}
\theta_{\rm med} =  \left\{
\begin{array}{ll}
1.2~{\rm arcsec~for~} S_{1.4~\rm GHz} < 1~{\rm mJy},\\
2(S_{1.4~\rm GHz})^{0.3}~{\rm arcsec~otherwise }.\\
\end{array}\right.\label{eqn:thmed2}
\end{equation}
Using this we can calculate the resolution bias correction factor to be applied to the source counts: $c = 1/[1 - h(>\Psi_{\rm max})]$. The correction factors calculated using the two different median size distributions (given by equations~\ref{eqn:thmed1} and \ref{eqn:thmed2}) are plotted as a function of flux density in Fig~\ref{figs:bias_correction}. We use the mean of these two correction factors to correct the source counts in this paper. The difference between these two correction factors is used to estimate the uncertainty in the source counts due to resolution bias; this is added in quadrature to the overall uncertainty in the source counts. Some of the brightest sources still have large correction factors ($\sim 1.1$) as they are located near the edge of the map where the noise is highest.

The resolution-bias corrected and completeness-corrected source count in each flux density bin is given by:
\begin{equation}
\frac{1}{A} \sum_{i=1}^{N} \frac{c}{P(S_i > 5\sigma)}
\end{equation}
where $A$ is the total area of the field, $N$ is the number of sources in the bin, $P(S_i > 5\sigma)$ is the probability of detecting a source with flux density $S_i$ in the field (given by Equation~\ref{eqn:prob}) and $c$ is the resolution bias correction.

\subsection{Sample variance}\label{section:cosmic_variance}

The influence of source clustering on radio source counts at 1.4~GHz was investigated by \citet{2013MNRAS.432.2625H}. They extracted a series of independent samples from the \citet{2008MNRAS.388.1335W} simulation and used this to present a method for estimating the uncertainty induced by sample variance on an arbitrary radio survey. Their analysis assumes that the noise is constant across the survey, which is not the case in our observations. The best r.m.s.\ noise in our image is 18~$\muup$Jy, while the r.m.s.\ noise towards the edge of the field used to calculate the source counts is 90~$\muup$Jy. Using Fig.~2 from \citet{2013MNRAS.432.2625H}, and converting to 610~MHz using a spectral index of 0.75, we expect the uncertainty due to sample variance on these observations to be $\approx$ 7~per cent (this remains relatively constant with flux density as the effective area covered by our observations increases as flux density increases). This uncertainty is included when calculating the overall uncertainty in our source counts.

\subsection{Other possible biases}

Statistical fluctuations due to thermal noise can alter the flux density of sources, causing them to be put into the `wrong' bins (Eddington Bias; \citealt{1913MNRAS..73..359E}). Due to the shape of the source counts this causes more sources to be scattered into a bin than out of it, and therefore introduces a positive bias. This effect is only significant near the detection limit of a survey. We follow \citet{2008MNRAS.383...75G} and make no correction for Eddington Bias, but note that this could cause the number of observed sources to be slightly too high in the fainter bins.

\subsection{Final source counts}

\begin{table}
\caption{610-MHz differential source counts. The counts are listed both before and after the resolution bias correction has been applied.}\label{tab:counts}
\smallskip
\centering
\begin{tabular}{ccccc}
\hline
Flux density & bin mid.$^{a}$ & d$N$/d$S$ uncorr.  & d$N$/d$S$ corr. & $\Delta$ d$N$/d$S$ \\
bin (mJy)    &  (mJy)         &(Jy$^{-1}$ sr$^{-1}$) & (Jy$^{-1}$ sr$^{-1}$) & (Jy$^{-1}$ sr$^{-1}$)\\
\hline
  0.08 -- 0.13 & 0.11 &  5.44$\times 10^{10}$ & 5.98$\times 10^{10}$ & 0.72$\times 10^{10}$\\
  0.13 -- 0.19 & 0.16 &  2.70$\times 10^{10}$ & 2.95$\times 10^{10}$ & 0.25$\times 10^{10}$\\
  0.19 -- 0.28 & 0.23 &  1.02$\times 10^{10}$ & 1.10$\times 10^{10}$ & 0.09$\times 10^{10}$\\
  0.28 -- 0.42 & 0.35 &  4.31$\times 10^{9}$ & 4.65$\times 10^{9}$ & 0.42$\times 10^{9}$\\
  0.42 -- 0.62 & 0.50 &  1.96$\times 10^{9}$ & 2.10$\times 10^{9}$ & 0.21$\times 10^{9}$\\
  0.62 -- 0.92 & 0.76 &  1.00$\times 10^{9}$ & 1.06$\times 10^{9}$ & 0.12$\times 10^{8}$\\
  0.92 -- 1.37 & 1.13 &  4.43$\times 10^{8}$ & 4.64$\times 10^{8}$ & 0.65$\times 10^{7}$\\
  1.37 -- 2.04 & 1.67 &  1.28$\times 10^{8}$ & 1.35$\times 10^{8}$ & 0.29$\times 10^{7}$\\
  2.04 -- 3.04 & 2.42 &  7.81$\times 10^{7}$ & 8.16$\times 10^{7}$ & 1.82$\times 10^{7}$\\
  3.04 -- 4.62 & 3.75 &  5.25$\times 10^{7}$ & 5.48$\times 10^{7}$ & 1.23$\times 10^{7}$\\
  4.52 -- 6.72 & 5.52 &  2.12$\times 10^{7}$ & 2.22$\times 10^{7}$ & 0.64$\times 10^{6}$\\
  6.72 -- 152.0 & 8.04 & 8.30$\times 10^{6}$ & 8.75$\times 10^{6}$ & 0.33$\times 10^{6}$\\
\hline
\end{tabular}
Notes:
(a) This is the mean flux density in the bin.
\end{table}

\begin{figure}
\centerline{\includegraphics[width=\columnwidth]{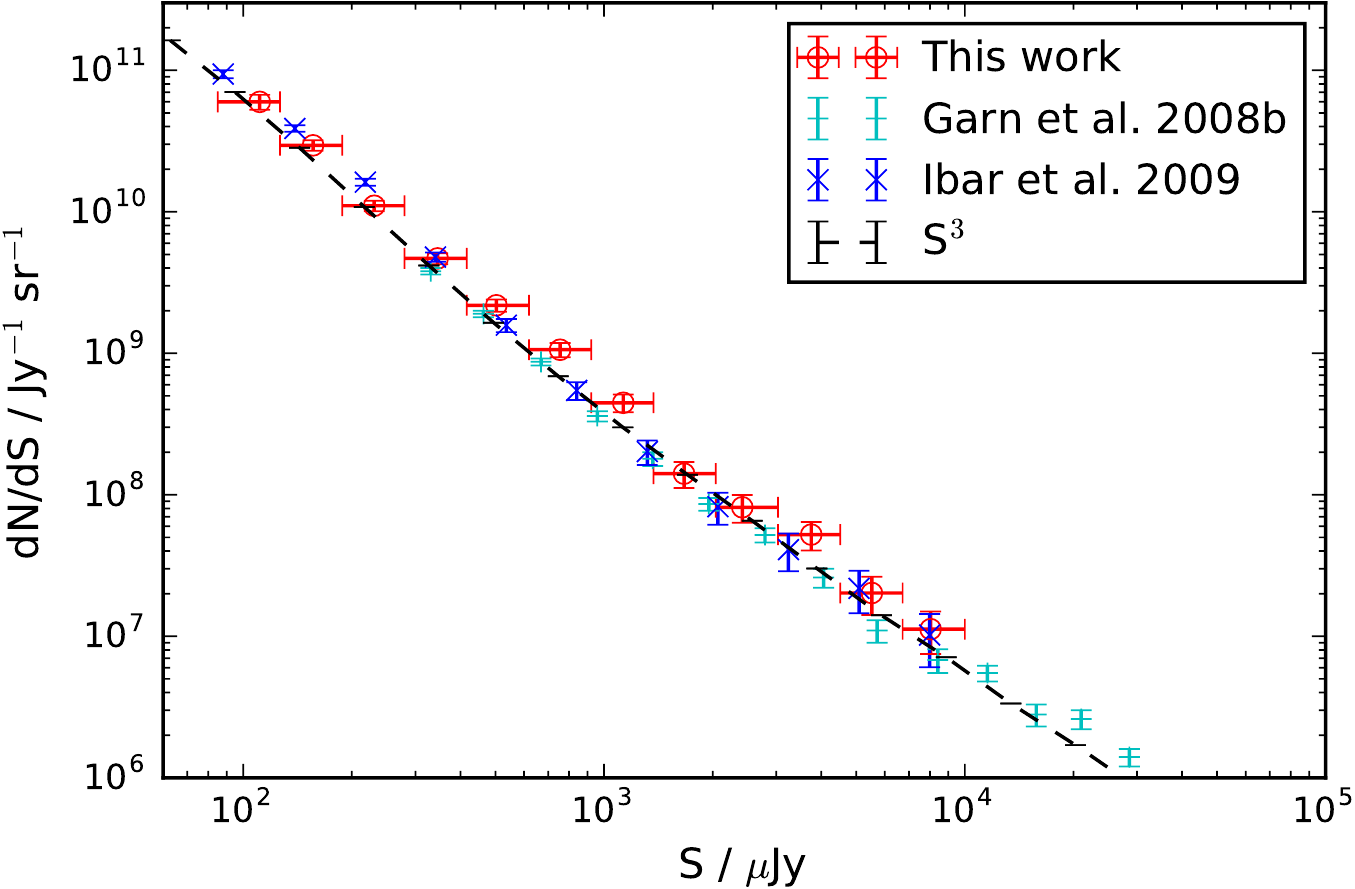}}
\caption{610-MHz source counts. The result from this work is shown, along with previous results from \citet{2008MNRAS.387.1037G} and \citet{2009MNRAS.397..281I}, as well as from the \citet{2008MNRAS.388.1335W} semi-analytic model (labelled S$^3$).}\label{fig:counts}
\end{figure}

\begin{figure*}
\centerline{\includegraphics[width=17cm]{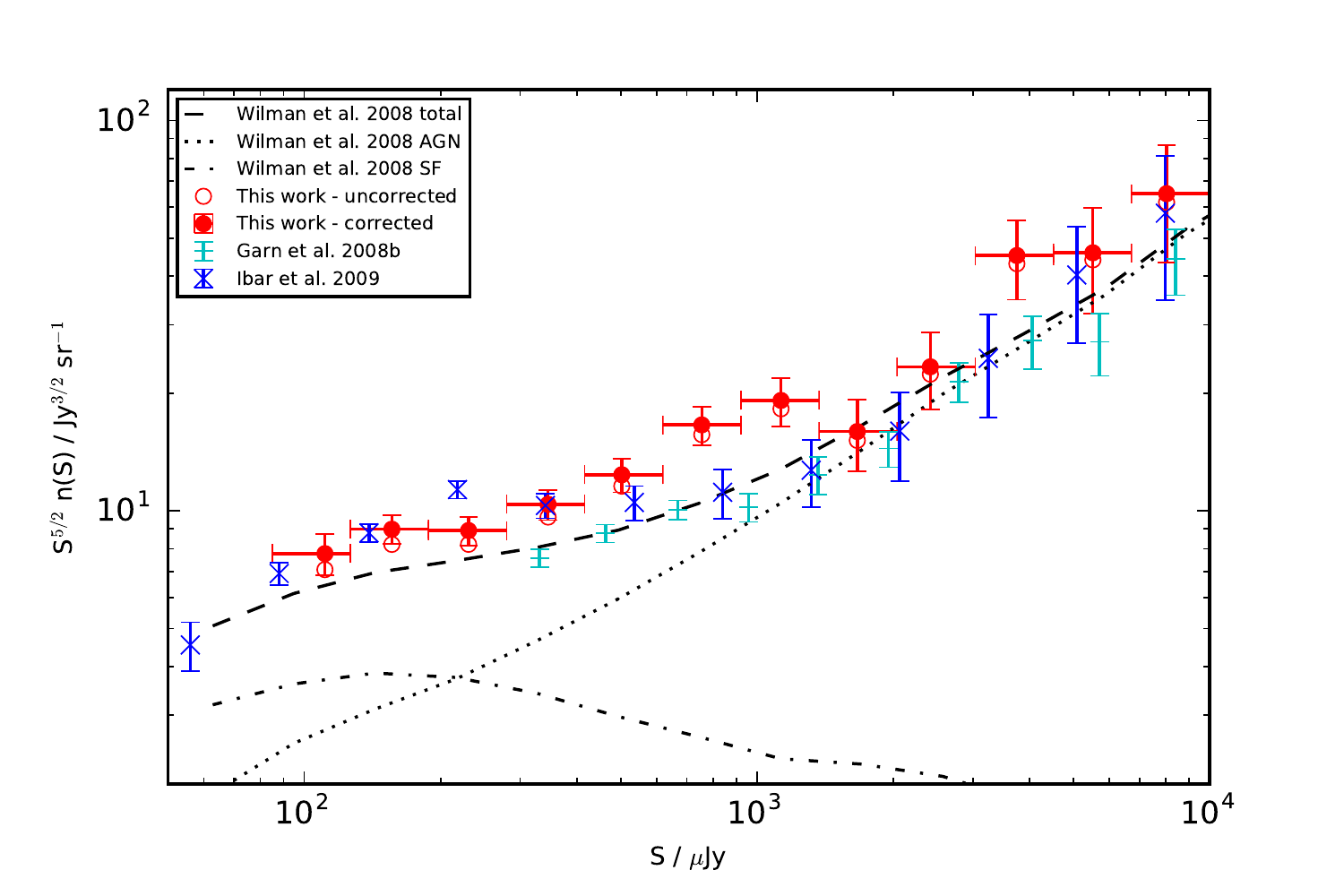}}
\caption{Euclidian normalised 610-MHz source counts. The filled circles show the source count corrected for resolution bias and the open circles show the uncorrected count. The horizontal error bars represent the bin width. Previous results from \citet{2008MNRAS.387.1037G} and \citet{2009MNRAS.397..281I} are shown, along with the \citet{2008MNRAS.388.1335W} semi-analytic model, which is split into AGN and star-forming sources.}\label{fig:counts_eucl}
\end{figure*}

The source counts are presented in Table~\ref{tab:counts} and are plotted in Fig.~\ref{fig:counts}. The Euclidean normalised counts are shown in Fig.~\ref{fig:counts_eucl}. The points are plotted at the mean flux density in each bin. The vertical error bars are the $\sqrt{n}$ Poisson uncertainties combined in quadrature with the resolution bias uncertainty (the difference between the two possible correction factors described in Section~\ref{section:bias}) and the 7~per cent uncertainty due to sample variance (see Section~\ref{section:cosmic_variance}). The horizontal error bars represent the bin widths. In Table~\ref{tab:counts} and Fig.~\ref{fig:counts_eucl} the source counts are given both before and after the correction for resolution bias has been applied; this correction increases the source counts by a small amount.

Previous 610-MHz GMRT source counts by \citet{2008MNRAS.387.1037G} and \citet{2009MNRAS.397..281I}, as well as the 610-MHz counts from the semi-analytic SKADS Simulated Skies (S$^3$) produced by \citet{2008MNRAS.388.1335W} are shown in Figs.~\ref{fig:counts} and \ref{fig:counts_eucl} for comparison. The \citeauthor{2008MNRAS.387.1037G} source counts are from areas of 5~deg$^2$ in the Lockman Hole field, 9~deg$^2$ in the ELAIS-N1 field \citep{2008MNRAS.383...75G} and 4~deg$^2$ in the \emph{Spitzer} extragalactic First Look Survey field \citep{2007MNRAS.376.1251G}. The \citet{2009MNRAS.397..281I} source counts are from 1~deg$^2$ in the Lockman Hole. In all three cases the error bars plotted are the $\sqrt{N}$ Poisson errors.

The source counts show good agreement with these previous results, except between $0.6 < S_{610~\rm MHz} / \rm mJy < 1.4$, where our observations show values of $S^{5/2} {\rm d}N/{\rm d}S$ which are higher by a factor of $\approx 1.5$ when compared to the other catalogues. \citet{2008MNRAS.387.1037G} compared the differential source counts from several different surveys at 610~MHz and found a difference of a factor of two in $S^{5/2} {\rm d}N/{\rm d}S$ below 1~mJy, which shows that this sort of discrepancy between source counts derived from different observations is not unusual. There is also significant scatter in different measurements of the differential source counts at 1.4~GHz, many of which do not agree with each other within their respective errors (see e.g.\ \citealt{2005AJ....130.1373H,2006MNRAS.371..963B,2007ASPC..380..189C,2008AJ....136.1889O}). As discussed in Section~\ref{section:cosmic_variance}, the expected uncertainty in these counts due to cosmic variance estimated from the \citeauthor{2013MNRAS.432.2625H} study is $\approx 7$~per cent at the flux limit, which is not large enough to account for the difference seen here. If these discrepancies are not due to source clustering, other possibilities include issues relating to calibration uncertainties, different methods for correcting for resolution bias (e.g. \citealt{2008ApJ...681.1129B}), or uncertainties in the primary beam correction and smearing effects (e.g. \citealt{2006ApJS..167..103F}). 

The \citet{2008MNRAS.388.1335W} simulation under-predicts the 610-MHz source counts observed by both our results and those by \citeauthor{2009MNRAS.397..281I} below approximately 0.5~mJy by a factor of 1.4. This could be because either the density of star-forming galaxies, or AGN (or both) are under-estimated in the simulation. However, as we have seen that differences of this order of magnitude between different surveys are relatively common, this difference could be due to instrumental effects as discussed above, rather than a failure in the simulation. (\citealt{2010MNRAS.404..532M} also present source count models at 610~MHz, but these do not go below a few mJy.)

\section{Spectral index distribution}\label{section:alpha}

The spectral index of a source $\alpha$, where $S \propto \nu^{- \alpha}$, can provide useful information about its nature. Star-forming sources generally have steep spectra, with spectral indices of $\alpha \approx 0.7$, due to synchrotron emission from supernovae \citep{1992ARA&A..30..575C}. AGN, however, can display a wide range of spectral indices, from rising through to very steep, depending on their structure and orientation with respect to the observer \citep{2010A&ARv..18....1D}. Classical double radio galaxies (e.g.\ Fanaroff and Riley type I and II sources, \citealt{1974MNRAS.167P..31F}) display powerful extended jets which produce steep-spectrum ($\alpha > 0.5$) synchrotron emission, while their cores have much flatter spectra ($\alpha < 0.5$) due to the superposition of many synchrotron self-absorbed spectra. The overall spectral index of a radio galaxy can inform us about the relative contributions of the extended structure and the core to the total emission observed. 

\citet{2013MNRAS.429.2080W} studied the spectral index distribution of sources selected from the 10C survey at 15.7 GHz and found a significant change with flux density; above $S_{15.7~\rm GHz} \sim 1$~mJy the sample was dominated by steep-spectrum sources, while at $S_{15.7~\rm GHz} \lesssim 1$~mJy sources with much flatter spectra dominated the population. This showed that the nature of the population was changing as flux density decreased. Using the recent deeper extension to the 10C survey (10C ultra-deep; \citealt{2016MNRAS.457.1496W}), along with the GMRT observations described in this paper, we are able to extend this study to fainter flux densities.

\subsection{Matching the catalogues}\label{section:matching}

The GMRT observations described in this paper cover the deepest part of the 10C ultra-deep observations in the AMI001 field. The 10C ultra-deep observations where made with the AMI Large Array at 15.7~GHz and have a lowest r.m.s.\ noise of 16~$\muup$Jy beam$^{-1}$. The AMI data (synthesised beam 30~arcsec) has a lower resolution than the 610-MHz GMRT data (synthesised beam 7~arcsec). The 10C ultra-deep catalogue contains 159 sources in the area covered by the GMRT observations. 

A match radius of 15~arcsec was used in the earlier work at higher flux densities \citep{2013MNRAS.429.2080W}, as this was found to maximise the number of real associations while mostly avoiding false matches. For consistency, the same match radius is used here. 133 out of the 159 10C sources have a match to the GMRT catalogue within 15~arcsec.

The difference in resolution between the 610-MHz and 15.7-GHz observations can cause problems when matching the two catalogues, as sources which are unresolved in the AMI data may be resolved into several components in the GMRT data. We therefore extracted 2.5~arcmin `postage stamp' images from the GMRT map at the position of each 10C source and examined them by eye. If there was only one GMRT source within 15~arcsec of the 10C source and that GMRT source was flagged as a point source by the source fitting algorithm then the flux density of that GMRT source was accepted as a match. If the GMRT source was extended, the GMRT map was convolved with a 30~arcsec Gaussian, to create a map with similar resolution to the AMI map (synthesised beam size 30~arcsec), and the flux density of the GMRT source was extracted from the smoothed map using the \textsc{aips} task \textsc{jmfit}. If the GMRT source was resolved into several components in the smoothed map, or \textsc{jmfit} did not converge (which was the case for a small number of sources), then the flux density of the source was extracted from the smoothed map by hand using the \textsc{aips} task \textsc{tvstat} down to the $2.5\sigma$-level.

\begin{figure}
\centerline{\includegraphics[width=\columnwidth]{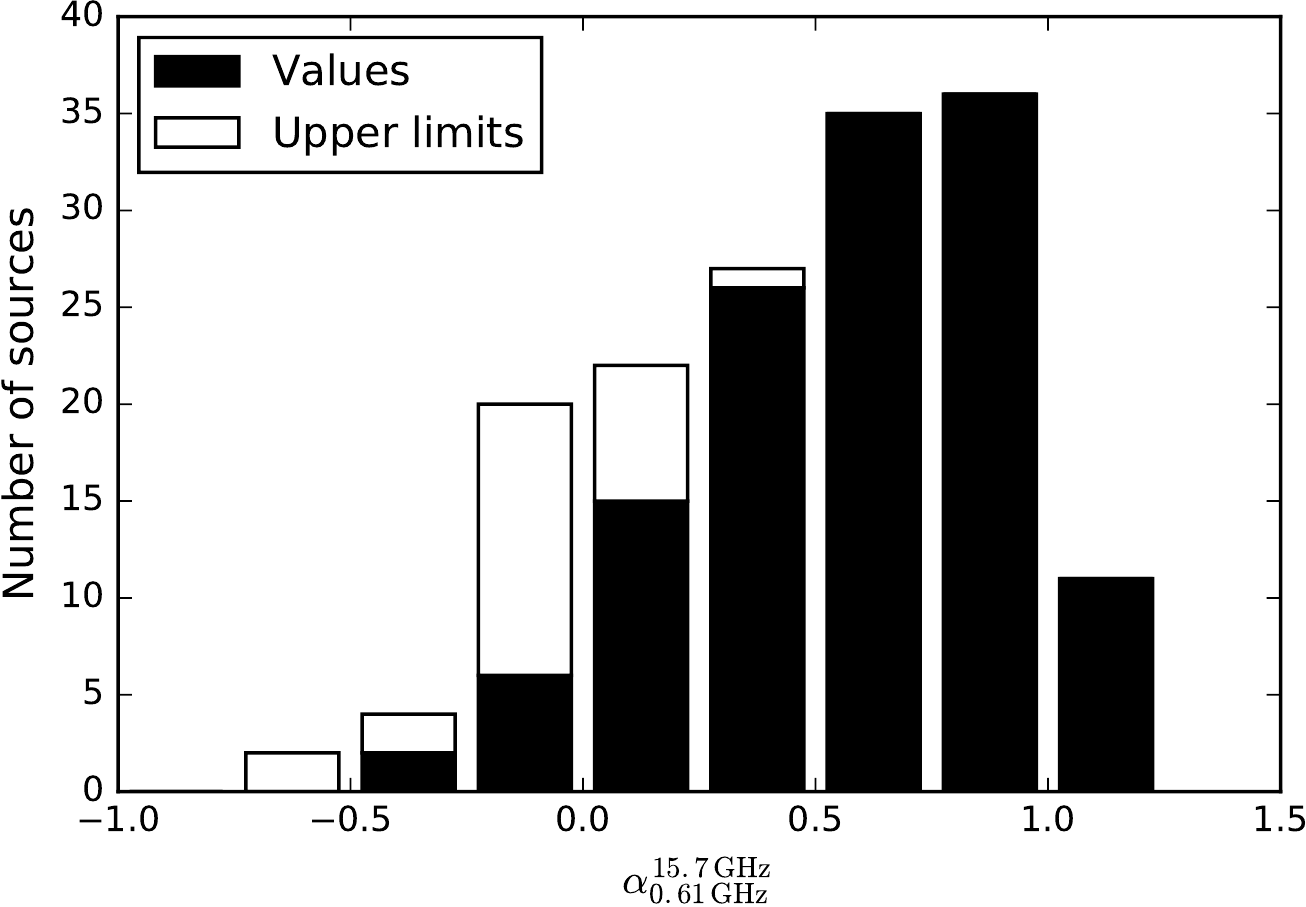}}
\caption{Distribution of spectral index between 610~MHz and 15.7~GHz ($\alpha$ where $S \propto \nu^{-\alpha}$) of 15.7-GHz selected sources in the AMI001 field. Values shown in white are upper limits, and could therefore move to the left.}\label{fig:alpha_hist}
\end{figure}

\begin{figure}
\centerline{\includegraphics[width=\columnwidth]{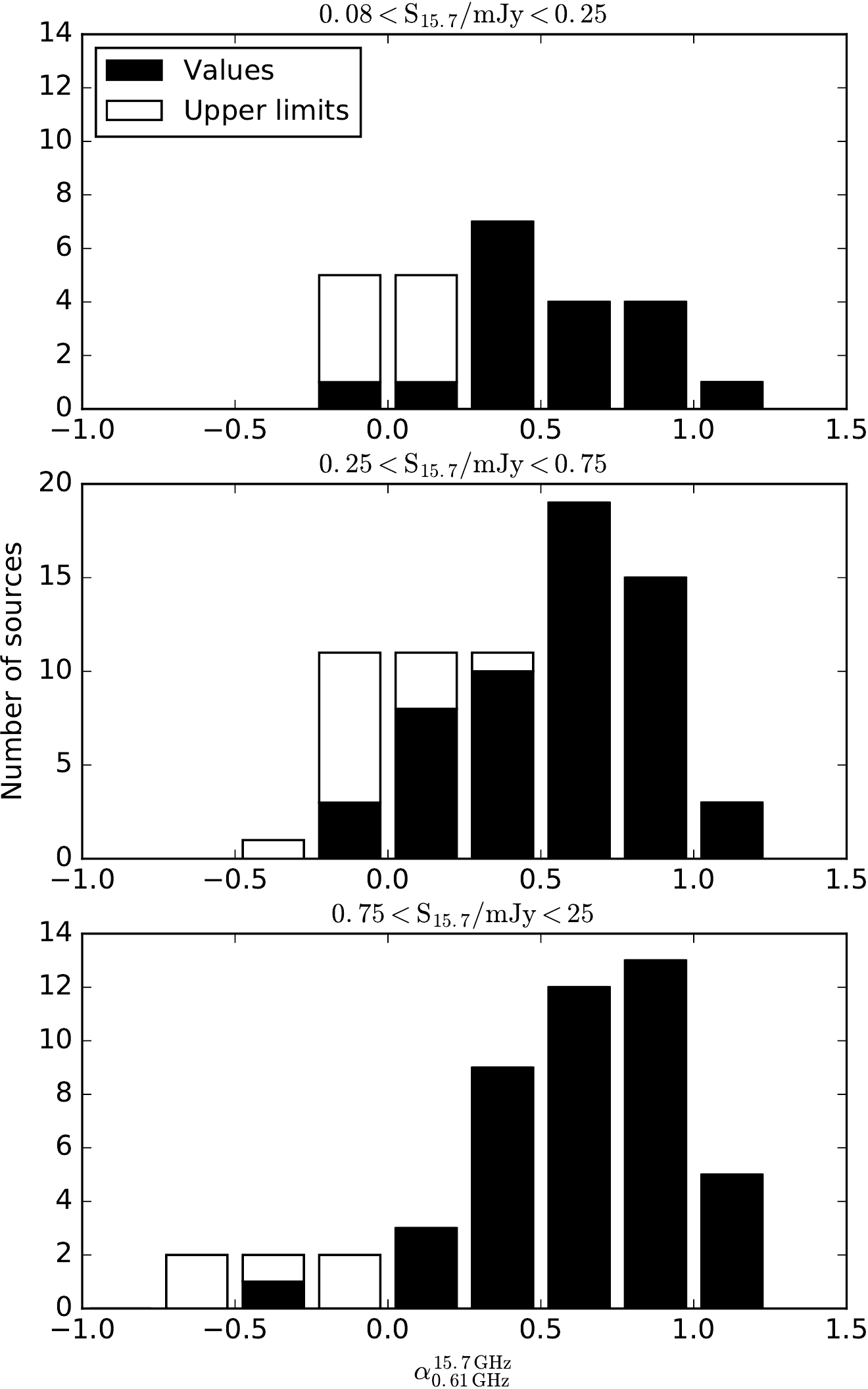}}
\caption{Distribution of spectral index between 610~MHz and 15.7~GHz ($\alpha$ where $S \propto \nu^{-\alpha}$) for 15.7-GHz selected sources in the AMI001 field in three different 15.7-GHz flux density bins. Values shown in white are upper limits, and could therefore move to the left.}\label{fig:alpha_bins_hist}
\end{figure}

26 10C ultra-deep sources do not have a match in the GMRT catalogue. The smoothed GMRT postage stamp images of these sources were examined by eye to see if there was a source visible which had fallen below the signal-to-noise threshold for inclusion in the GMRT catalogue, nothing was detected at the positions of the 26 sources. Therefore, an upper limit on the 610-MHz flux density of these sources was taken to be three times the local noise in the smoothed postage stamp image (estimated with the \textsc{aips} task \textsc{imean}).

In order to increase the sample size and reduce the possible effects of cosmic variance, spectral indices are also calculated for sources in the second 10C ultra-deep field, the Lockman Hole field. The 15.7-GHz observations in this field are less deep than those in the AMI001 field, with a lowest r.m.s.\ noise of 21~$\muup$Jy~beam$^{-1}$ as opposed to 16~$\muup$Jy~beam$^{-1}$, and cover a smaller area. To calculate spectral indices, the 10C ultra-deep catalogue in the Lockman Hole field is matched to the catalogue from a deep 1.4-GHz image of the region observed with the Westerbork Synthesis Radio Telescope (WSRT) \citep{2012rsri.confE..22G}. The image has an r.m.s.\ noise in the centre of $11~\muup$Jy and a synthesised beam size of $11 \times 9$~arcsec$^2$. This WSRT catalogue was matched to the original 10C catalogue by \citet{2013MNRAS.429.2080W} using a match radius of 15~arcsec, and we follow the same procedure here when matching it to the 10C ultra-deep catalogue. The resolutions of the two catalogues are similar so there are fewer potential pitfalls when matching these catalogues.

There are 137 sources in the 10C ultra-deep Lockman Hole field (this includes 58 sources in the original 10C catalogue), 131 of which have a match in the WSRT catalogue within 15 arcsec. Several sources are flagged as having multiple components in the WSRT catalogue; these sources were examined by eye to determine whether they are also resolved into multiple components in the 10C ultra-deep catalogue. This was found to be the case in most instances so the flux densities of the separate components were used; for the remaining sources the flux densities of the 1.4-GHz components were combined. WSRT images of the six unmatched sources were examined by eye and in no case was a source visible below the detection threshold; an upper limit of three times the local noise was therefore placed on the 1.4-GHz flux densities of these sources.

\subsection{The spectral index distribution}

\begin{figure}
\centerline{\includegraphics[width=\columnwidth]{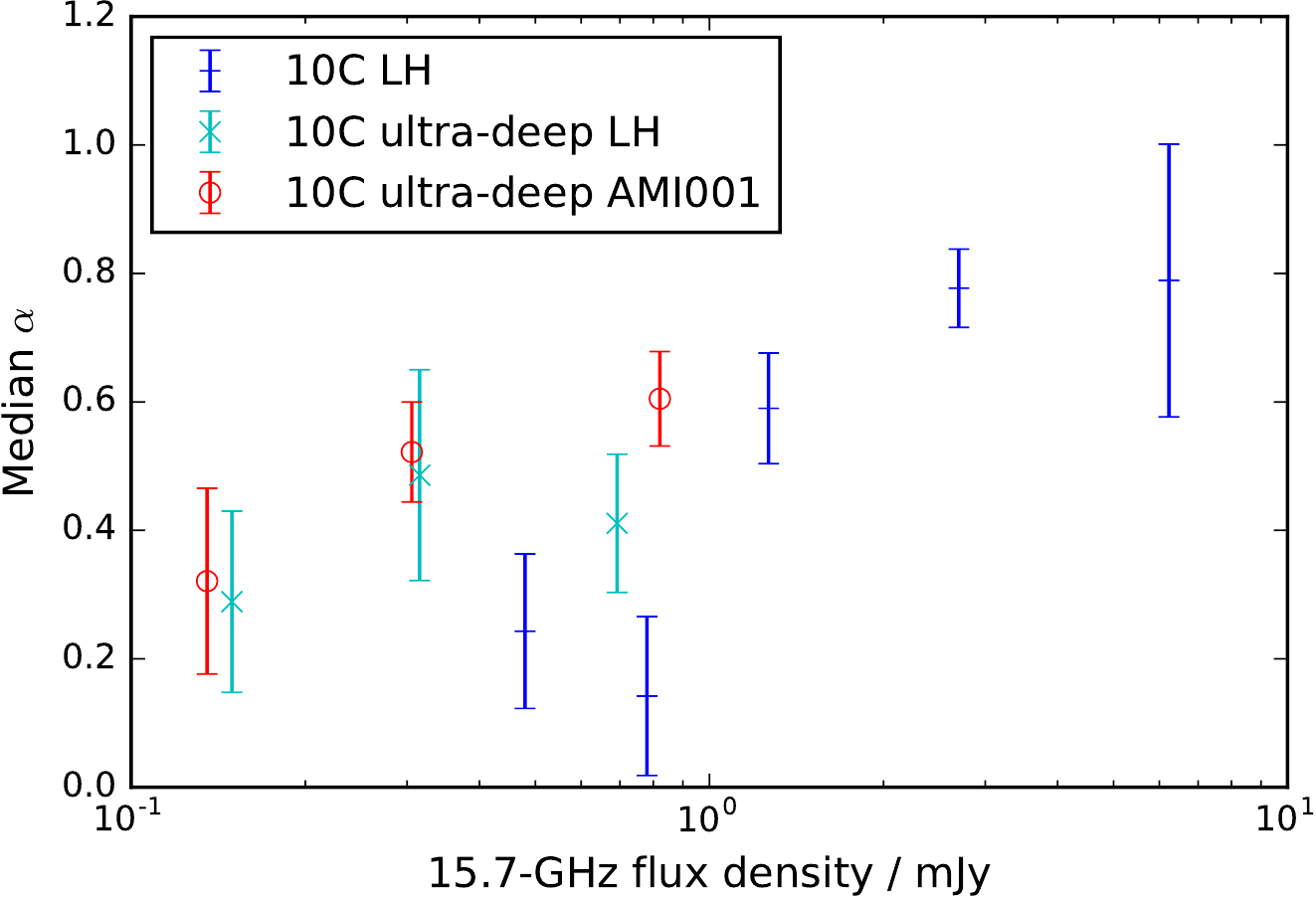}}
\caption{Median spectral index ($\alpha$ where $S \propto \nu^{-\alpha}$) in different 15.7-GHz flux density bins for sources in the 10C ultra-deep surveys in the Lockman Hole and AMI001 fields, and for 10C sources from \citet{2013MNRAS.429.2080W}. For the two Lockman Hole samples spectral indices are between 1.4 and 15.7~GHz while for the AMI001 field they are between 610~MHz and 15.7~GHz. Medians are calculated using survival analysis to take into account upper limits.}\label{fig:alpha}
\end{figure}

\begin{figure}
\centerline{\includegraphics[width=\columnwidth]{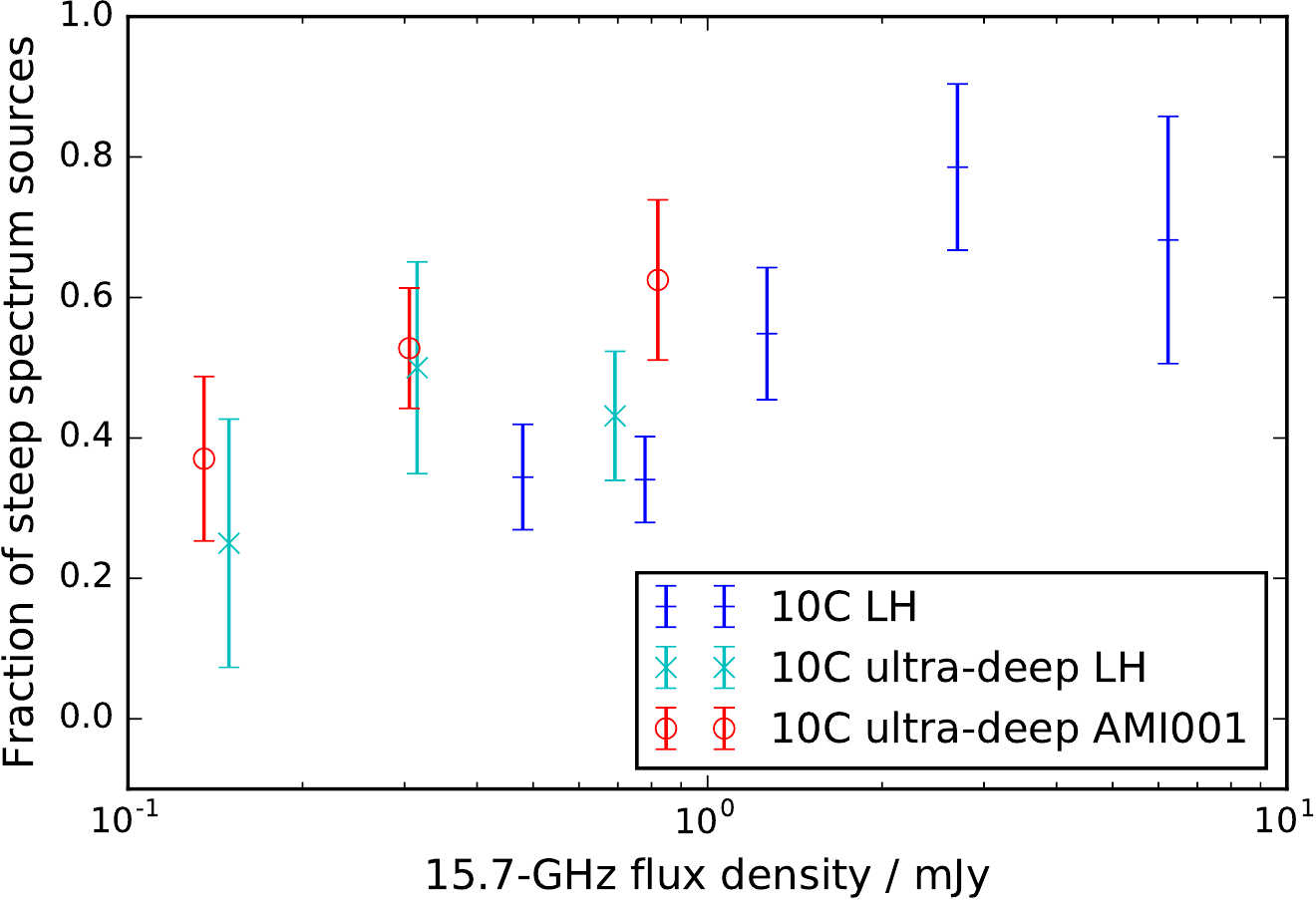}}
\caption{Fraction of steep-spectrum ($\alpha > 0.5$) sources in different 15.7-GHz flux density bins for 10C ultra-deep sources in the Lockman Hole and AMI001 fields, and for 10C sources from \citet{2013MNRAS.429.2080W}. For the two Lockman Hole samples $\alpha^{15.7}_{1.4}$ is used while for the AMI001 field $\alpha^{15.7}_{0.61}$ is used.}\label{fig:alpha_frac}
\end{figure}

The spectral index distribution of all 159 sources in the 15.7-GHz selected sample in the AMI001 field is shown in Fig.~\ref{fig:alpha_hist}. Upper limits are plotted for the 26 sources without a counterpart in the GMRT catalogue and are shown in white, these values could therefore move to the left. The median spectral index for the sample is $0.52 \pm 0.05$. Throughout this section, the median values are calculated using the \textsc{asurv} Rev. 1.2 package which takes into account upper limits by implementing the survival analysis methods presented in \citet{1985ApJ...293..192F}.

The sample is split into three separate flux density bins in Fig.~\ref{fig:alpha_bins_hist}, which shows that the highest flux-density bin ($0.75 < S/{\rm mJy} < 25$) has a higher proportion of steep spectrum sources than the two lower flux density bins, in which there are increasing numbers of flat-spectrum sources. This result is shown more clearly in Fig.~\ref{fig:alpha}, which shows the median spectral index in several 15.7-GHz flux density bins. The 10C ultra-deep Lockman Hole sample and the original 10C results from \citet{2013MNRAS.429.2080W} are also shown in this figure. Note that for the two samples in the Lockman Hole the spectral indices are between 1.4~GHz and 15.7~GHz ($\alpha^{15.7}_{1.4}$), while for the AMI001 field they are between 610~MHz and 15.7~GHz ($\alpha^{15.7}_{0.61}$). Fig.~\ref{fig:alpha} shows that the median spectral index decreases (indicating that the source spectra become flatter) as the flux density decreases, and that the median spectral index remains less than 0.5 down to $S_{15.7~\rm GHz} = 0.1$~mJy. There is, however, significant scatter in the median spectral indices from the three samples, particularly around $S_{15.7~\rm GHz} = 1$~mJy. Fig.~\ref{fig:alpha_frac} shows the fraction of steep-spectrum ($\alpha > 0.5$) sources in the same flux density bins for the three samples and indicates that the proportion of steep-spectrum sources decreases as flux density decreases.

These results suggest that star-forming galaxies are not making a significant contribution to the 15.7-GHz source population at flux densities down to $S_{15.7~\rm GHz} \sim 0.1~\rm mJy$, as these sources typically have steeper spectra (with $\alpha \approx 0.75$). This is consistent with the conclusions of \citet{2016MNRAS.457.1496W}, who used the 10C ultra-deep data to calculate the 15.7-GHz source counts down to 0.1~mJy and found no evidence for the emergence of a significant new population of sources, such as star-forming galaxies. It is in contrast to the predictions made by the SKADS Simulated Skies \citep{2008MNRAS.388.1335W}, which predicts that 20 per cent of the sources with $0.1 < S_{18~\rm GHz} / \rm mJy < 0.3$ should be star-forming galaxies. It seems that star-forming galaxies are not contributing to the high-frequency source population at the levels predicted by the simulation. 

Assuming they have steep spectra, the population of star-forming galaxies which are responsible for the inflection in the 1.4-GHz source counts observed at 1~mJy (e.g.\ \citealt{2004NewAR..48.1173J,2008MNRAS.386.1695S}) should begin to contribute to the 15.7-GHz source counts at around $S_{15.7~\rm GHz} = 0.1$~mJy, the limit of this study. Extending this work to fainter flux densities should therefore allow us to detect these star-forming galaxies, and see whether or not they contribute to the high-frequency sky at the levels expected at fainter flux densities. We refer the reader to Section~7 of \citet{2016arXiv160900499P} for a recent review of the future prospects for studying the faint radio sky with upcoming surveys.

\section{Conclusions}\label{section:conclusions}

We have made new 610-MHz observations with the GMRT of 0.84~deg$^2$ with a lowest r.m.s.\ noise of 18~$\muup$Jy beam$^{-1}$. The differential source counts derived from these observations between $0.1 < S_{610~\rm MHz}/ \rm mJy < 10$ are in agreement with other work within the expected scatter between different surveys. 
The SKA Simulated Skies underestimate the Euclidean normalised source counts by a factor of 1.4 at $S_{610~\rm MHz} < 1$~mJy; however as this difference is a similar order of magnitude to the scatter found between different surveys, it could be due to cosmic variance and instrumental effects rather than a failure in the model.

These deep 610~MHz observations enable us to investigate the spectral index distribution of an unique sample of radio sources selected at 15.7~GHz. We find that this population continues to be dominated by flat spectrum sources down to $S_{15.7~\rm GHz} = 0.1$~mJy, with a median spectral index of $0.32 \pm 0.14$ between $0.08 < S_{15.7~\rm GHz} / \rm mJy < 0.2$. This suggests that star-forming galaxies make no significant contribution to this population. This is in agreement with a recent study of the source counts of this sample \citep{2016MNRAS.457.1496W} and provides further evidence that the SKA Simulated Skies do not accurately model the radio source population at high frequencies.

\section*{Acknowledgements}

We thank the anonymous referee for their careful reading of this paper. We thank the GMRT staff who made these observations possible. The GMRT is run by the National Centre for Radio Astrophysics of the Tata Institute of Fundamental Research. IHW thanks the Science and Technologies Facilities Council for a studentship. IHW and MJJ acknowledge support from the Square Kilometre Array South Africa. IHW thanks the South African Astronomical Observatory, where some of this work was carried out.

%
%

\setlength{\labelwidth}{0pt}

\bsp

\label{lastpage}
\end{document}